\documentclass[
aps,prb,superscriptaddress,twocolumn,showpacs,preprintnumbers,amsmath,amssymb, 10pt
]{revtex4-2}

\usepackage{graphicx, color}
\usepackage{float}
\usepackage{dcolumn}
\usepackage{bm}
\usepackage{enumerate}
\usepackage{hyperref}

\begin{document}

\title{
Impact of the out-of-plane conductivity on spin transport evaluation \\
in a van der Waals material
}

\author{Ryoya~Nakamura}
\affiliation{Department of Physics, Graduate School of Science, The University of Osaka, Toyonaka 560-0043, Japan}
\author{Futo~Tokuda}
\affiliation{Department of Physics, Graduate School of Science, The University of Osaka, Toyonaka 560-0043, Japan}
\author{Yoshinobu~Ono}
\affiliation{Department of Physics, Graduate School of Science, The University of Osaka, Toyonaka 560-0043, Japan}
\author{Nan~Jiang}
\affiliation{Department of Physics, Graduate School of Science, The University of Osaka, Toyonaka 560-0043, Japan}
\affiliation{Center for Spintronics Research Network, The University of Osaka, Toyonaka 560-8531, Japan}
\affiliation{Institute for Open and Transdisciplinary Research Initiatives, The University of Osaka, Suita 565-0871, Japan}
\author{Hideaki~Sakai}
\affiliation{Department of Physics, Graduate School of Science, The University of Osaka, Toyonaka 560-0043, Japan}
\author{Masayuki~Ochi}
\affiliation{Department of Physics, Graduate School of Science, The University of Osaka, Toyonaka 560-0043, Japan}
\affiliation{Forefront Research Center, The University of Osaka, Toyonaka, Osaka 560-0043, Japan}
\author{Hiroaki~Ishizuka}
\affiliation{Department of Physics, Institute of Science Tokyo, Meguro, Tokyo 152-8551, Japan}
\author{Yasuhiro~Niimi}
\email{niimi@phys.sci.osaka-u.ac.jp}
\affiliation{Department of Physics, Graduate School of Science, The University of Osaka, Toyonaka 560-0043, Japan}
\affiliation{Center for Spintronics Research Network, The University of Osaka, Toyonaka 560-8531, Japan}
\affiliation{Institute for Open and Transdisciplinary Research Initiatives, The University of Osaka, Suita 565-0871, Japan}

\begin{abstract}
Layered materials are promising candidates for spintronic applications due to their unique electronic structures and spin transport properties.  However, the strong anisotropic conductivity inherent in these materials complicates the quantitative evaluation of spin Hall conductivity and spin diffusion length.
In this work, we present a comprehensive study of spin transport in a transition metal dichalcogenide PtTe$_2$ by combining a three-dimensional finite element model with nonlocal spin valve structures.
We developed a theoretical model that treats an anisotropic spin diffusion in the same way as the conventional isotropic model, enabling the extraction of spin diffusion lengths along both the in-plane and out-of-plane directions.
Our analysis revealed that the conventional isotropic assumption tends to overestimate some values, particularly for the out-of-plane spin diffusion length and spin Hall conductivity.
These findings provide new insight into anisotropic spin diffusion and spin-charge conversions in layered materials and emphasize the importance of accounting for anisotropic conductivity in the design of spintronic devices.

\end{abstract}

\maketitle

\section{\label{section1}Introduction}
The spin Hall effect (SHE) is a key concept in spintronics, where charge current is converted into spin current in materials with strong spin-orbit interactions (SOIs)~\cite{Sinova2015}.
This conversion is quantified by the spin Hall conductivity~$\sigma_{\mathrm{SH}}$, expressed as $\bm{J}_{\mathrm{s}} = \sigma_{\mathrm{SH}}(\bm{E} \times \bm{s})$, where $\bm{J}_{\mathrm{s}}$ represents the spin current, $\bm{E}$ is the electric field, and $\bm{s}$ denotes the spin direction.
For magnetic random-access memory applications, the efficiency of spin-orbit-torque-induced magnetization reversal depends crucially on the spin Hall conductivity~\cite{Manchon2019}. Hence, a precise evaluation of this property is essential.

In spintronics, heavy metals such as Pt~\cite{Saitoh2006, Vila2007, Sagasta2016, Liu2011, Morota2011}, Ta~\cite{Liu2012, Kim2012, Morota2011}, and W~\cite{Pai2012, Hao2015} have long been the central focus because of their large spin Hall angles.
Recently, layered materials grown by molecular beam epitaxy (MBE) and mechanically exfoliated van der Waals (vdW) materials have attracted significant attention.
For instance, a giant spin Hall angle resulting from spin-momentum locking has been observed in topological insulators~\cite{Mellnik2014, Fan2014}.
The sign change of SHE has been achieved by the magnetic-order parameter reversal in Mn$_3$Sn~\cite{Kimata2019} or by ferroelectric switching in GeTe~\cite{Varotto2021}.
Furthermore, multidirectional SHE with reduced crystal symmetry has been observed in transition metal dichalcogenides (TMDs) such as 1$T'$-WTe$_2$~\cite{MacNeill2017, Zhao2020, Camosi2022, Kao2022}, 1$T'$-MoTe$_2$~\cite{Stiehl2019, Safeer2019, Ontoso2023, Ontoso2023_2}, and 1$T$-TaS$_2$ below the charge density wave transition temperature~\cite{Chi2024}.

Among various TMDs, 1$T$-PtTe$_2$ is a promising candidate for spintronic applications due to its unique electronic properties.
PtTe$_2$ is classified as a type-II Dirac semimetal~\cite{Yan2017, Politano2018, Fei2018, Jiang2020}. It has the highest conductivity among TMDs~\cite{Pavlosiuk2018} and exhibits a strong SOI as indicated by weak antilocalization measurements~\cite{Hao2018, Song2022}.
Recent experiments have demonstrated its large spin Hall conductivity~\cite{Xu2020} and efficient terahertz emission~\cite{Yadav2024, Li2025}, suggesting topological surface state contributions.
In addition, a giant spin Hall conductivity due to the out-of-plane anti-damping torque~\cite{Wang2024, Wang2024_2, Yang2025} has been demonstrated in PtTe$_2$/WTe$_2$/CoFeB trilayer.
These findings highlight PtTe$_2$ as an ideal platform to investigate the SHE.

However, one crucial issue must be addressed to accurately evaluate the spin Hall conductivity $\sigma_{\mathrm{SH}}$ in layered materials. 
That is the strong anisotropy inherent in layered materials.
To our knowledge, all SHE-studies in layered materials with a spin diffusion length $\lambda_{\mathrm{s}}$ shorter than $\approx 100$~nm have assumed the isotropic resistivity when calculating spin transport properties. 
In other words, the strong anisotropy in layered materials has been neglected so far. Therefore, it is important to develop evaluation methods that account for the anisotropy.

In this work, we quantitatively evaluated 
the spin diffusion length and the spin Hall conductivity of PtTe$_2$ 
using a three-dimensional (3D) finite element method (FEM). 
We first measured the inverse spin Hall effect (ISHE) of PtTe$_2$ nanowire 
and spin accumulation signals with and without the nanowire in 
nonlocal spin valve (NLSV) devices. 
To calculate the spin diffusion length and spin Hall conductivity in an anisotropic system, we developed a theoretical model where an anisotropic spin diffusion can be treated in the same way as the conventional isotropic model.
Based on this model, we obtained the spin diffusion length along both the in-plane and out-of-plane directions.
We then evaluated the spin Hall conductivity and found that it decreases when 
the anisotropy is taken into account.
The results demonstrate the limitation of the conventional isotropic model and reveal that neglecting anisotropy in layered materials can lead to overestimations of the out-of-plane spin diffusion length and spin Hall conductivity.
Additionally, we observed a crossover from intrinsic SHE in the lower-conductivity regime to extrinsic SHE in the higher-conductivity regime.
These findings provide new insight into the anisotropic spin diffusion and spin-charge conversion in layered materials.

\section{Anisotropic model\label{sec:2}}
In NLSV measurements, the spin transport channel is typically regarded as quasi-one-dimensional (1D), and the 1D spin diffusion model developed by Takahashi and Maekawa~\cite{Takahashi2003, Takahashi2008} has been widely used. This model assumes that the spin diffusion takes place only along the out-of-plane ($z$-axis) direction 
in an SHE material. 
However, when the spin diffusion length becomes comparable to or longer than the thickness of the SHE material, the 1D model tends to underestimate the spin diffusion length; therefore, a 3D spin diffusion model based on an extension of the Valet-Fert formalism is necessary for an accurate evaluation~\cite{Niimi2012}.
Furthermore, the 1D model could not be applicable in systems with strong anisotropy, such as layered materials, where the out-of-plane conductivity $\sigma_{\perp}$ is much smaller than the in-plane conductivity $\sigma_{\parallel}$, i.e., $\sigma_{\perp} \ll \sigma_{\parallel}$. In such a case, spin currents preferentially diffuse along the in-plane direction, and thus the 1D model does not accurately describe the spin transport behavior.
To address these issues, we perform numerical calculations.

In the following, we assume that the spin polarization is collinear and 
oriented along the $x$-axis.
The system exhibits a uniaxial conductivity anisotropy, with $\sigma_{\parallel}$ and $\sigma_{\perp}$ denoting the in-plane and out-of-plane conductivities, respectively. 
We treat the spin relaxation time $\tau_{\mathrm{sf}}$ as an isotropic parameter [see Appendix~\ref{appendix:A} for details of the assumption].
Under these assumptions, the spin diffusion equations in a steady state read
\begin{subequations}
\begin{align}
  &\nabla\cdot
  \begin{pmatrix}
  \sigma_{\parallel}\partial_x(\mu_{\uparrow}+\mu_{\downarrow}) \\
  \sigma_{\parallel}\partial_y(\mu_{\uparrow}+\mu_{\downarrow})
  + \sigma_{\mathrm{SH}}\partial_z(\mu_{\uparrow}-\mu_{\downarrow})\\
  \sigma_{\perp}\partial_z(\mu_{\uparrow}+\mu_{\downarrow})
  - \sigma_{\mathrm{SH}}\partial_y(\mu_{\uparrow}-\mu_{\downarrow})\\
  \end{pmatrix}
  = 0,
  \label{eq:1a} \\
  \frac{1}{2e}&\nabla\cdot
  \begin{pmatrix}
  \sigma_{\parallel}\partial_x(\mu_{\uparrow}-\mu_{\downarrow}) \\
  \sigma_{\parallel}\partial_y(\mu_{\uparrow}-\mu_{\downarrow})
  + \sigma_{\mathrm{SH}}\partial_z(\mu_{\uparrow}+\mu_{\downarrow})\\
  \sigma_{\perp}\partial_z(\mu_{\uparrow}-\mu_{\downarrow})
  - \sigma_{\mathrm{SH}}\partial_y(\mu_{\uparrow}+\mu_{\downarrow})\notag\\
  \end{pmatrix} \\
  &= \frac{e\rho(\epsilon_{\mathrm{F}})}{\tau_{\mathrm{sf}}}(\mu_{\uparrow}-\mu_{\downarrow}),
  \label{eq:1b}
\end{align}
\end{subequations}
where $e$, $\mu_{\uparrow, \downarrow}$, $\rho(\epsilon_{\mathrm{F}})$, and $\tau_{\mathrm{sf}}$ are, respectively, the electric charge, the electrochemical potentials of up- and down-spin electrons, the density of states at the Fermi energy, and the spin relaxation time [see Appendix~\ref{appendix:B}].
For numerical simulations, it is convenient to introduce a renormalized axis $x = x'$, $y = y'$, $z = \sqrt{\sigma_{\perp}/\sigma_{\parallel}}\,z'$~\cite{Johnson2009, Piraud2013, Ijpdea2014, Vafek2023}.
Using this renormalization, the system can be regarded as isotropic with the conductivity $\sigma_{\parallel}$.
Using the renormalized axis, Eqs.~\eqref{eq:1a} and ~\eqref{eq:1b} can be written as follows:
\begin{subequations}
\begin{align}
   &\sigma_{\parallel}{\nabla'}^2(\mu_{\uparrow}+\mu_{\downarrow}) + \sqrt{\frac{\sigma_{\parallel}}{\sigma_{\perp}}}\sigma_{\mathrm{SH}}(\partial_{y'}\partial_{z'}-\partial_{z'}\partial_{y'})(\mu_{\uparrow}-\mu_{\downarrow}) \notag\\
   &= 0, \label{eq:2a} \\
   &\sigma_{\parallel}{\nabla'}^2(\mu_{\uparrow}-\mu_{\downarrow}) + \sqrt{\frac{\sigma_{\parallel}}{\sigma_{\perp}}}\sigma_{\mathrm{SH}}(\partial_{y'}\partial_{z'}-\partial_{z'}\partial_{y'})(\mu_{\uparrow}+\mu_{\downarrow}) \notag\\
   &= \frac{\sigma_{\parallel}}{D_{\parallel}\tau_{\mathrm{sf}}}(\mu_{\uparrow}-\mu_{\downarrow})
   = \frac{\sigma_{\perp}}{D_{\perp}\tau_{\mathrm{sf}}}(\mu_{\uparrow}-\mu_{\downarrow}), \label{eq:2b} 
\end{align}
\end{subequations}
where $D_{\parallel}$ and $D_{\perp}$ are the diffusion constants for in-plane and out-of-plane directions, respectively.
Note that, from Einstein relation, $D_{\parallel,\perp}/\sigma_{\parallel,\perp}=1/(2e^2\rho(\epsilon_{\mathrm{F}}))$.
The anisotropy of spin diffusion length, $\mathrm{\lambda}_{\mathrm{s}}^{\parallel,\perp} = \sqrt{D_{\parallel,\perp}\tau_{\mathrm{sf}}}$, as 
$\mathrm{\lambda}_{\mathrm{s}}^{\parallel}/\mathrm{\lambda}_{\mathrm{s}}^{\perp}=\sqrt{D_{\parallel}/D_{\perp}}=\sqrt{\sigma_{\parallel}/\sigma_{\perp}}$.
Equations~\eqref{eq:2a} and \eqref{eq:2b} enable us to calculate $\lambda_{\mathrm{s}}^{\parallel}$ and $\sqrt{\sigma_{\parallel}/\sigma_{\perp}}\,\sigma_{\mathrm{SH}}$ as free parameters as in the case of the isotropic model. 
We can obtain $\lambda_{\mathrm{s}}^{\perp}$ and $\sigma_{\mathrm{SH}}$ by transforming the obtained results back to the original axis.

\section{\label{section3}Experimental details}

\begin{figure}
  \begin{center}
    \includegraphics[width = \linewidth]{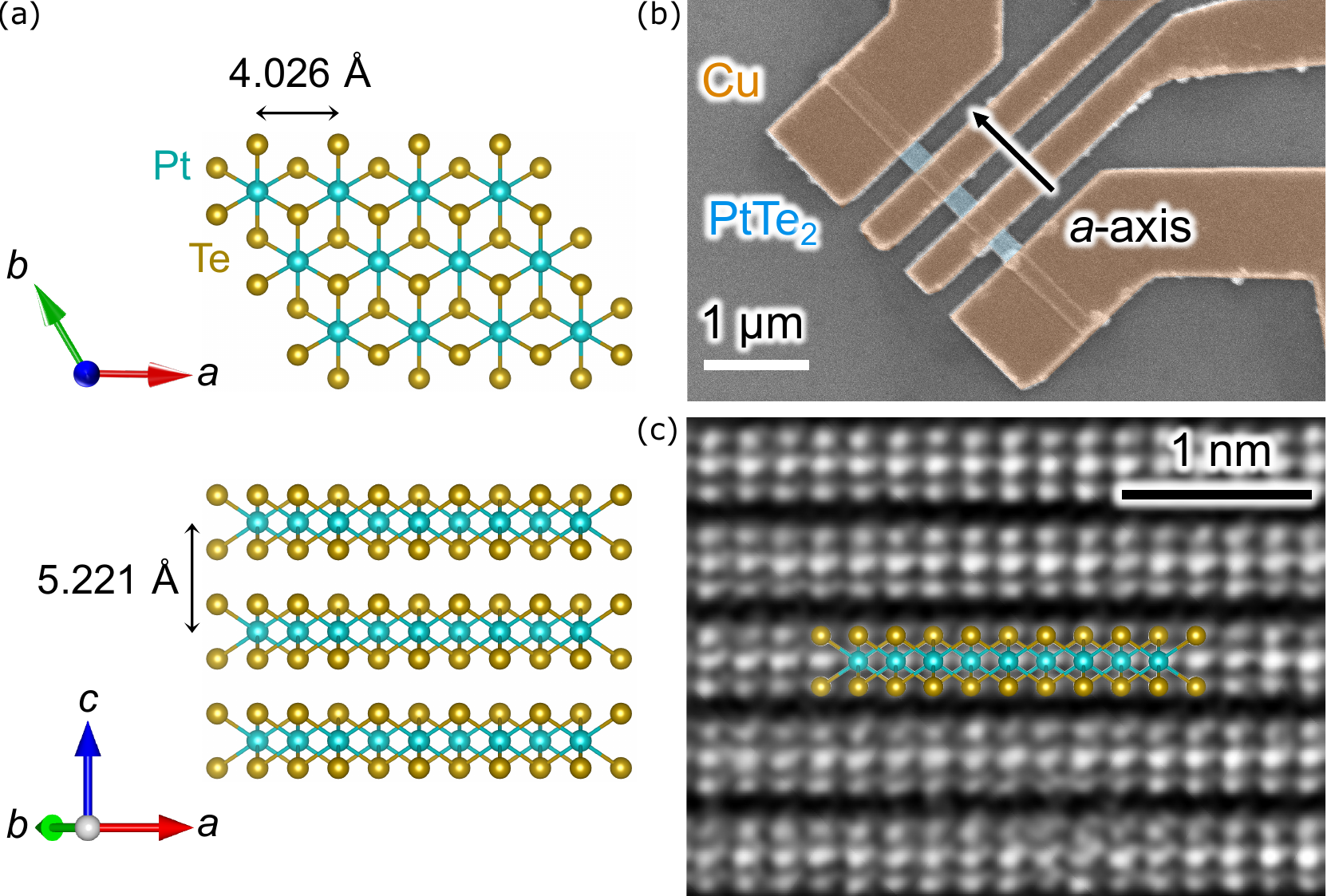}
    \caption{
        (a) Schematic of the crystal structure of PtTe$_2$ (space group $P\overline{3}m1$, No. 164), illustrated using VESTA~\cite{VESTA2011}.
        Pt and Te atoms are shown as blue and yellow spheres, respectively.
        (b) False-colored SEM image of a typical device used for four-terminal measurements. 
        (c) Cross-sectional TEM image of a PtTe$_2$ thin film along its longer direction.
    }
    \label{fig:1}
  \end{center}
\end{figure}

PtTe$_2$ has the CdI$_2$-type trigonal (1$T$) structure with the $P\overline{3}m1$ space group (No.~164) as illustrated in Fig.~\ref{fig:1}(a).
PtTe$_2$ flakes were exfoliated from a commercial bulk crystal (2D Semiconductors) using a standard mechanical exfoliation process and subsequently transferred onto a 
SiO$_2$(285 nm)/Si substrate.
Notably, some of these flakes naturally form nanowire-like structures, as shown in the inset of Fig.~\ref{fig:1}(b).
This shape implies that the flakes cleave preferentially along the $a$-axis, similar to other TMDs such as ReS$_2$ and PtSe$_2$~\cite{Liu2015, Sun2020}.
This was confirmed by cross-sectional transmission electron microscopy (TEM), as shown in Fig.~\ref{fig:1}(c).

The NLSV structure was fabricated using the conventional lift-off method.
Py wires and Cu bridge were patterned by electron beam lithography onto the substrate coated with polymethyl-methacrylate (PMMA).
Py wires were first deposited adjacent to PtTe$_2$ flakes using an electron beam evaporator.
The width and thickness of the Py wires are 100 and 30~nm, respectively. Before the Cu bridge deposition, we carried out an Ar ion beam milling for 5 seconds to remove the residual resist from the surfaces of the Py wires and PtTe$_2$ flakes, which etched the PtTe$_2$ surface by approximately 1~nm.
The device was then transferred to another chamber without breaking the vacuum, and the Cu bridge was deposited using a Joule heating evaporator with a 99.9999\% purity source.
Both the width and thickness of the Cu bridge are 100~nm.
We also fabricated the NLSV devices without PtTe$_2$ flakes on the same substrate for reference.
For the resistivity measurements of thin films, we deposited four-terminal Cu electrodes on PtTe$_2$ flakes using the same process.
For the bulk crystal, gold wires were attached to the sample using silver paste.
Although PtTe$_2$ is known to be air-stable, all the fabrication processes except for electron beam lithography and wire bonding were performed in a glovebox filled with Ar gas to maintain clean interfaces.
All the electrical transport measurements were performed using an ac lock-in amplifier and a $^4$He flow cryostat. We confirmed the transparency of the Cu/Py and Cu/PtTe$_2$ interfaces for all the devices by measuring their interfacial resistance [see Appendix~\ref{appendix:C} for details of measurements].
The magnetic field was applied to the device using an electromagnet within the range of $\pm5000$~Oe.
The width $w$ and thickness $t$ of PtTe$_2$ flakes were determined using a scanning electron microscopy (SEM) and atomic force microscopy, respectively, after the transport measurements were completed.
The device parameters used for the spin transport measurements are summarized in Appendix~\ref{appendix:D}.

There are two advantages to use NLSV for quantitative evaluation.
First, since only the pure spin current is injected into the material, contributions from the local charge current can be ignored~\cite{Stiehl2019_2, Shi2025}, allowing us to calculate the spin accumulation distribution.
Second, the lack of contact with a ferromagnet eliminates magnetic proximity effects, which are known to enhance $\sigma_{\mathrm{SH}}$ in Pt and PtSe$_2$~\cite{Wang2016, Mudgal2023}.
In addition, a recent work has shown that PtTe$_2$ can also modify the magnetic properties of the adjacent ferromagnet Fe$_3$GaTe$_2$~\cite{Guo2025}

\section{\label{section4}Results}
\begin{figure}
  \begin{center}
    \includegraphics[width = \linewidth]{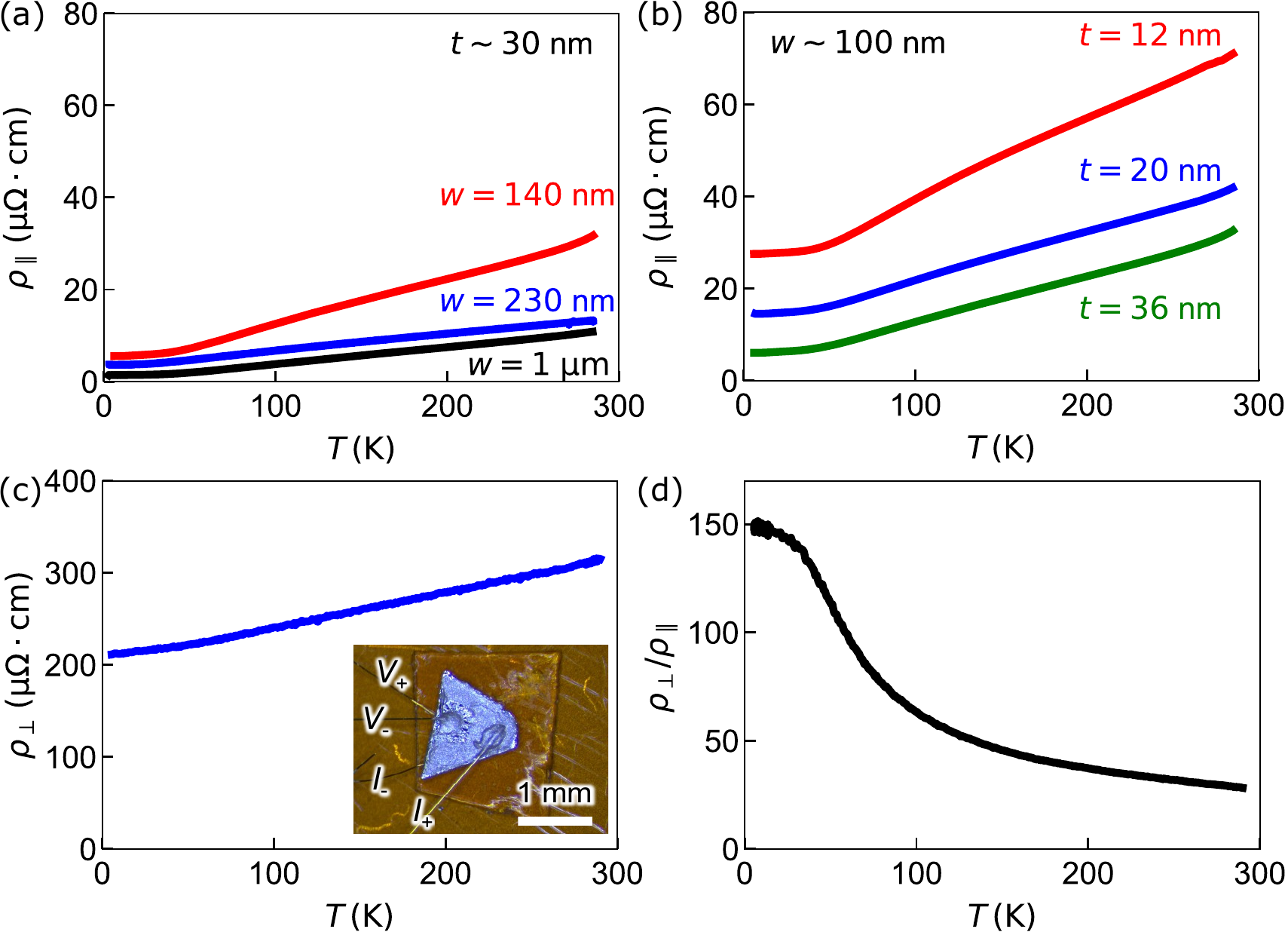}
    \caption{
      Temperature dependence of (a) $\rho_{\parallel}$ in a PtTe$_2$ thin film with thickness $t \approx 30$~nm and different widths, 
      (b) $\rho_{\parallel}$ in a thin film with width $w \approx 100$~nm and different thicknesses, 
      (c) $\rho_{\perp}$ in a bulk PtTe$_2$ (inset: optical microscope image of the sample), and 
      (d) the anisotropy ratio $\rho_{\perp}/\rho_{\parallel}$, with $\rho_{\perp}$ from the bulk crystal and $\rho_{\parallel}$ from a thin film ($t = 30$~nm, $w = 1$~$\mathrm{\mu m}$).
    }
    \label{fig:2}
  \end{center}
\end{figure}

In order to evaluate the in-plane resistivity $\rho_{\parallel}$ of PtTe$_{2}$, 
we first conducted four-terminal measurements on PtTe$_2$ thin films.
Figure~\ref{fig:2}(a) shows the temperature dependence of $\rho_{\parallel}$ in PtTe$_2$ thin films with a thickness $t$ of approximately 30~nm. 
When the width $w$ of PtTe$_2$ is relatively large (e.g., 1~$\mathrm{\mu m}$), $\rho_{\parallel}$ remains low and consistent with bulk values~\cite{Pavlosiuk2018}.
However, $\rho_{\parallel}$ increases with decreasing $w$. 
This width dependence suggests that lateral confinement enhances electron scattering, resulting in a higher resistivity.
In addition to this width dependence, we also observed a clear thickness dependence of $\rho_{\parallel}$.
As shown in Fig.~\ref{fig:2}(b), $\rho_{\parallel}$ for narrow wires ($w \sim 100$~nm) also increases with reducing the film thickness $t$.
This indicates that the vertical confinement (i.e., the reduction of $t$) further enhances scattering, leading to an additional increase in resistivity.
Such a thickness dependence has also been reported in PtTe$_2$ films fabricated by chemical vapor deposition (CVD) and MBE~\cite{Xu2020, Wang2024}.
By using the Fuchs-Sondheimer model, we confirmed that $\rho_{\parallel}$ in PtTe$_2$ thin films is subject to a size effect [see Appendix~\ref{appendix:E} for details].
$\rho_{\parallel}$ of the 30~nm thick film is approximately seven times larger at low temperatures and three times larger at room temperature compared to the bulk.
In spin transport measurements, thinner and narrower wires enhance both the SHE and NLSV signals. 
We used films with $t \approx 30$~nm and $w \approx 100$~nm for subsequent spin transport measurements.
In order to evaluate the spin diffusion length and spin Hall conductivity of PtTe$_2$ films, 
if the device structure allows, we measure the resistivity of PtTe$_{2}$ using a four-terminal method; otherwise, we use $\rho_{\parallel}$ from samples with similar $t$ and $w$.

Next, we measured the out-of-plane resistivity $\rho_{\perp}$, which is essential for evaluating the anisotropic spin diffusion.
One method for measuring the out-of-plane resistivity of layered materials is to 
dry-transfer a thin film onto pre-patterned electrodes and place top electrodes in order to sandwich the film. 
We initially applied this vertical contact method, but the high aspect ratio of the film caused non-uniform current distribution, leading to negative and unreliable resistivity values [see Appendix~\ref{appendix:F}].
A previous study on 1$T$-TaS$_2$~\cite{Martino2020} points out the issues with this method.
In that material, out-of-plane resistivity measurements using the vertical contact method on a thin film provide much higher values of $\rho_{\perp}$ ($700~\mathrm{m}\Omega\cdot\mathrm{cm}$)~\cite{Svetin2017}, compared to $1.5~\mathrm{m}\Omega\cdot\mathrm{cm}$ from focused ion beam processing and FEM simulations~\cite{Martino2020}.
Thus, in this work, we measured the out-of-plane resistivity of bulk crystals, as shown in the inset of Fig.~\ref{fig:2}(c).
For this measurement, current electrodes were fabricated to cover a wide area of the crystal surfaces for uniform current injection. 
Small voltage electrodes were then placed in the area enclosed by the current electrodes [see Appendix~\ref{appendix:G} for details of measurements].
Figure~\ref{fig:2}(c) shows that $\rho_{\perp}$ was 200-300~$\mathrm{\mu}\Omega\cdot\mathrm{cm}$.
As a result, the anisotropy ratio $\rho_{\perp}/\rho_{\parallel}$ for a thin film with $t = 30$~nm and $w = 1$~$\mathrm{\mu m}$ exceeded 100 at low temperatures and was approximately 30 at room temperature, as shown in Fig.~\ref{fig:2}(d).
This behavior is consistent with previous reports on Dirac electron systems~\cite{Cao2017, Novak2019}, where the strong anisotropy arises from the low in-plane resistivity.
As pointed out by Martino $et~al$.~\cite{Martino2020}, in general, $\rho_{\perp}$ in layered materials depends weakly on the material, typically around $1~\mathrm{m\Omega \cdot cm}$~\cite{Edman1998, Martino2020, Martino2021, Su2025, Cao2017, Novak2019, Wang2017, Kim2018, Heine2021, Vedeneev2013}. 
In contrast, the in-plane resistivity $\rho_{\parallel}$ varies significantly depending on the material, leading to a wide range of anisotropy ratios $\rho_{\perp}/\rho_{\parallel}$.
For example, the ratio $\rho_{\perp}/\rho_{\parallel}$ is approximately $100$ in Graphite~\cite{Edman1998}, $4$ in 1$T$-TaS$_2$~\cite{Martino2020}, $10$ in NbS$_2$~\cite{Martino2021}, $10$ in MoTe$_2$~\cite{Su2025}, $400$ in IrTe$_2$~\cite{Cao2017}, $50$ in ZrSiS~\cite{Novak2019}, $15$ in Fe$_3$GeTe$_2$~\cite{Wang2017, Kim2018}, $40$ in YBa$_2$Cu$_3$O$_{7-\delta}~$\cite{Heine2021}, and $3$ in FeSe~\cite{Vedeneev2013}.
The universality for $\rho_{\perp}$ strongly suggests that $\rho_{\perp}$ is determined by a common physical mechanism (interlayer hopping or tunneling) which is independent of the in-plane scattering mechanisms (surface scattering or defects).
In addition, the out-of-plane resistivity is generally less susceptible to size effects~\cite{Choi2012, Zheng2017, Mariani2025}, which notably affect the in-plane resistivity $\rho_{\parallel}$, as discussed above.
Based on these considerations, in the subsequent analysis, we calculated the spin diffusion length and spin Hall conductivity, primarily assuming that the PtTe$_2$ thin films exhibit the same $\rho_{\perp}$ as the bulk. However, considering a worst-case scenario, we also evaluated these parameters assuming that $\rho_{\perp}$ is ten times larger than the bulk value.

\begin{figure}
  \begin{center}
    \includegraphics[width = \linewidth]{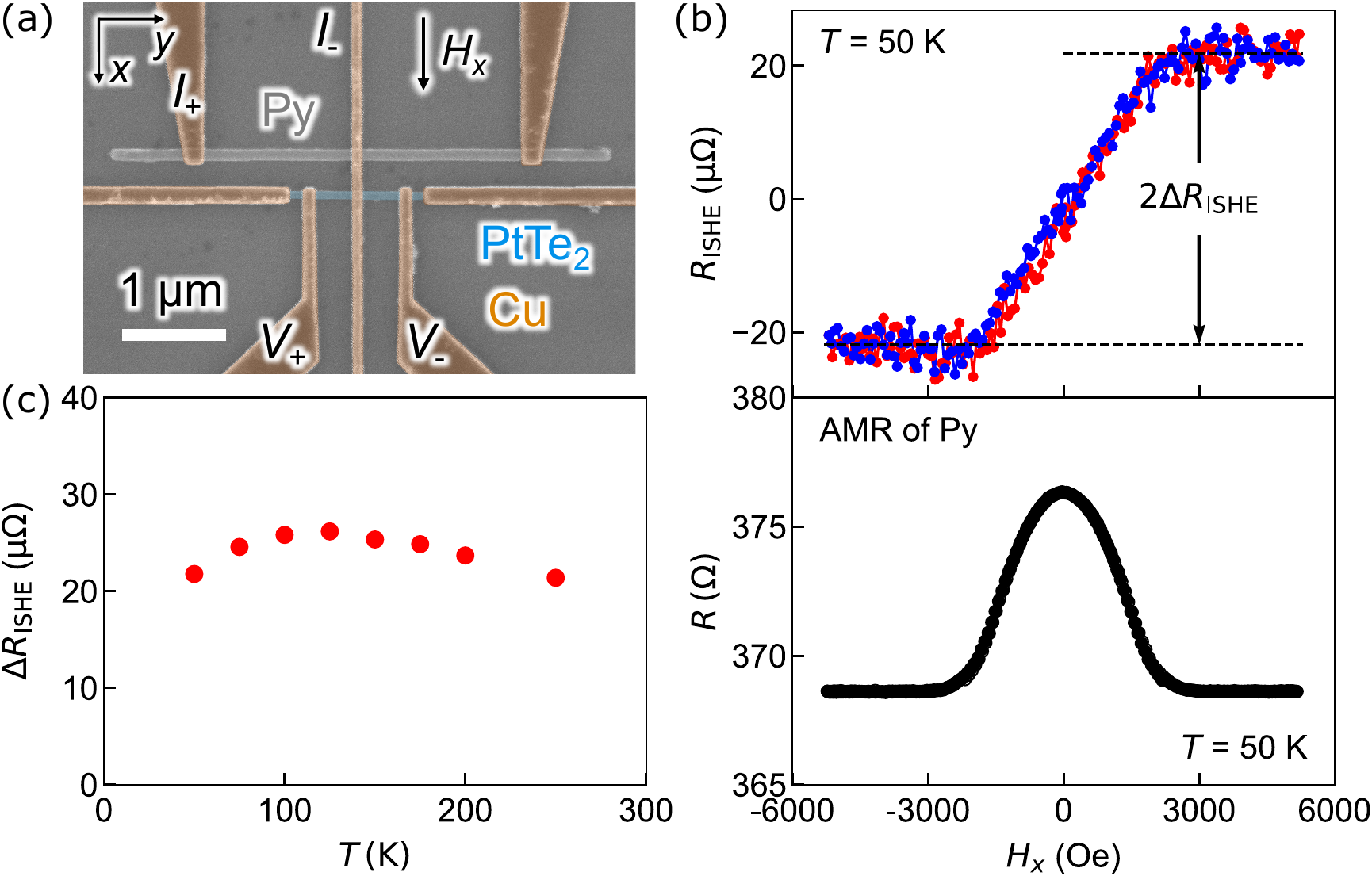}
    \caption{
      (a) False-colored SEM image of a typical device used for ISHE measurements.
      The arrow indicates the positive direction of the magnetic field, and the $x$ and $y$ axes define the in-plane coordinate system.
      (b) Top panel: ISHE resistance $R_{\mathrm{ISHE}}$ of PtTe$_2$ measured at 50~K for Sample D.
      Red and blue circles represent forward and backward magnetic field sweeps, respectively.
      Background signals have been subtracted from the raw data.
      The ISHE resistance amplitude $\Delta R_{\mathrm{ISHE}}$ is defined as the difference 
      between the average values of $R_{\mathrm{ISHE}}$ above $+3000$~Oe and below $-3000$~Oe, as indicated by the dashed lines.
      Bottom panel: Typical AMR signal of Py, showing the saturation of magnetization above 3000~Oe for $H_{x}$ applied along the hard axis.
      (c) Temperature dependence of the ISHE resistance amplitude $\Delta R_{\mathrm{ISHE}}$ for Sample D.
    }
    \label{fig:3}
  \end{center}
\end{figure}

We then measured the ISHE in PtTe$_2$ using a spin transport device 
as shown in Fig.~\ref{fig:3}(a).
By flowing a current $I$ from a Py wire to a Cu channel, nonequilibrium spin accumulation is generated at the Py/Cu interface.
Because this spin accumulation relaxes along both sides of the Cu channel, a spin current can flow within the spin diffusion length of Cu ($\approx 1~\mathrm{\mu m}$ at low temperatures).
The diffusive spin current is partly absorbed by a PtTe$_2$ flake.
In this PtTe$_2$ flake, the ISHE takes place: the spin current is converted into a charge current, resulting in a voltage drop ($V_+ - V_-$) along the flake when the spin direction is perpendicular to the wire direction.
The voltage drop ($V_+ - V_-$) normalized to the injected current $I$ is 
defined as the ISHE resistance $R_{\mathrm{ISHE}}$.
By switching the orientation of the magnetic field, the opposite $R_{\mathrm{ISHE}}$ is obtained since the Py magnetization is inverted as well as the orientation of the spin polarization.
In this configuration, the magnetic field direction aligns with the hard axis of the Py wire, and thus the magnetization of Py saturates above 3000~Oe, as indicated by the anisotropic magnetoresistance (AMR) effect in the bottom panel of Fig.~\ref{fig:3}(b).
The difference of the two $R_{\mathrm{ISHE}}$ values is twice the ISHE signal: 2$\Delta R_{\mathrm{ISHE}}$. 
In the top panel of Fig.~\ref{fig:3}(b), we show an ISHE resistance $R_{\mathrm{ISHE}}$ measured at 50~K.
When $H_{x}>3000$~Oe (or $< -3000$~Oe), $R_{\mathrm{ISHE}}$ is fully saturated.
It shows a clear positive ISHE signal, which is consistent with previous studies~\cite{Xu2020, Wang2024}.
Figure~\ref{fig:3}(c) shows the temperature dependence of $\Delta R_{\mathrm{ISHE}}$, which exhibits a maximum at $T \approx 120$~K. 
In the following analysis, we calculate the spin Hall conductivity $\sigma_{\mathrm{SH}}$ 
as a function of the conductivity as well as temperature using $\Delta R_{\mathrm{ISHE}}$. 

\begin{figure}
  \begin{center}
    \includegraphics[width = \linewidth]{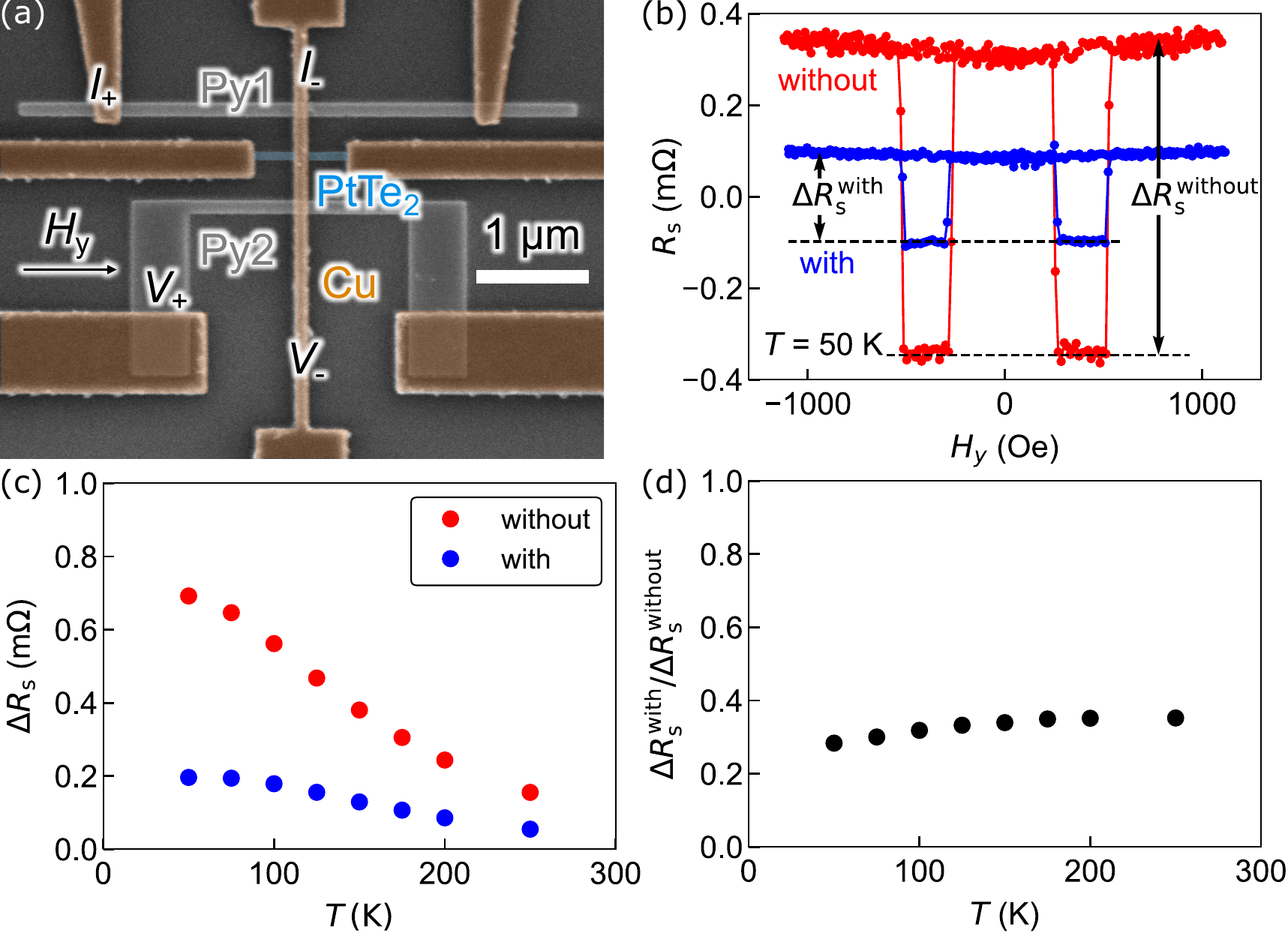}
    \caption{
        (a) False-colored SEM image of a typical device used for NLSV signal measurements with a PtTe$_2$ nanowire.
        The arrow represents the positive direction of the magnetic field.
        (b) NLSV signals without (red) and with (blue) the PtTe$_2$ middle wire measured at $T = 50$~K for Sample A.
        (c) Temperature dependence of the spin signal amplitudes $R_{\mathrm{s}}^{\mathrm{without}}$ (red) and $R_{\mathrm{s}}^{\mathrm{with}}$ (blue).
        (d) Temperature dependence of the amplitude ratio $R_{\mathrm{s}}^{\mathrm{with}}/R_{\mathrm{s}}^{\mathrm{without}}$.
        }
    \label{fig:4}
  \end{center}
\end{figure}

To estimate the spin diffusion length of PtTe$_2$ as well as to evaluate the amount of the spin current absorbed into PtTe$_2$, we performed NLSV measurements~\cite{Niimi2015} with and without the PtTe$_{2}$ nanowire.
For this purpose, we modified the device structure by adding the second Py nanowire, as shown in Fig.~\ref{fig:4}(a). 
We note that in this case the four-terminal resistivity cannot be measured 
because of the space of the device and thus the resistivity value obtained 
from a PtTe$_2$ flake with a similar thickness and width has been used for the analysis.
A magnetic field in this configuration is applied along the easy direction of the Py wires ($H_{y}$).
Similar to the previous configuration, the injected spin current diffuses along the Cu channel and is partially absorbed by the PtTe$_2$ flake due to its strong SOI.
The nonequilibrium spin accumulation is detected as a nonlocal voltage ($V_+ - V_-$) using the second Py wire.
By normalizing the measured voltage to the applied current $I$, the nonlocal resistance $R_{\mathrm{s}}^{\mathrm{with}}$ is defined.
$R_{\mathrm{s}}^{\mathrm{with}}$ depends on the magnetization of the two Py wires, i.e., whether they are in a parallel or antiparallel state [see the blue line in Fig.~\ref{fig:4}(b)], and the corresponding spin signal is defined as $\Delta R_{\mathrm{s}}^{\mathrm{with}}$.
Compared to the device with PtTe$_2$, the reference device without PtTe$_2$ exhibits a much larger spin signal $\Delta R_{\mathrm{s}}^{\mathrm{without}}$ [Fig.~\ref{fig:4}(b), red line].
Figure~\ref{fig:4}(c) shows the temperature dependence of the spin signals both $R_{\mathrm{s}}^{\mathrm{without}}$ and $R_{\mathrm{s}}^{\mathrm{with}}$.
Figure~\ref{fig:4}(d) presents the temperature dependence of the ratio of these spin signals, $\Delta R_{\mathrm{s}}^{\mathrm{with}}/\Delta R_{\mathrm{s}}^{\mathrm{without}}$.
In the following analysis, we calculate the spin diffusion length of PtTe$_2$ from $\Delta R_{\mathrm{s}}^{\mathrm{with}}/\Delta R_{\mathrm{s}}^{\mathrm{without}}$.

\section{\label{section5}Discussion}

\begin{figure}[htbp]
  \begin{center}
    \includegraphics[width = \linewidth]{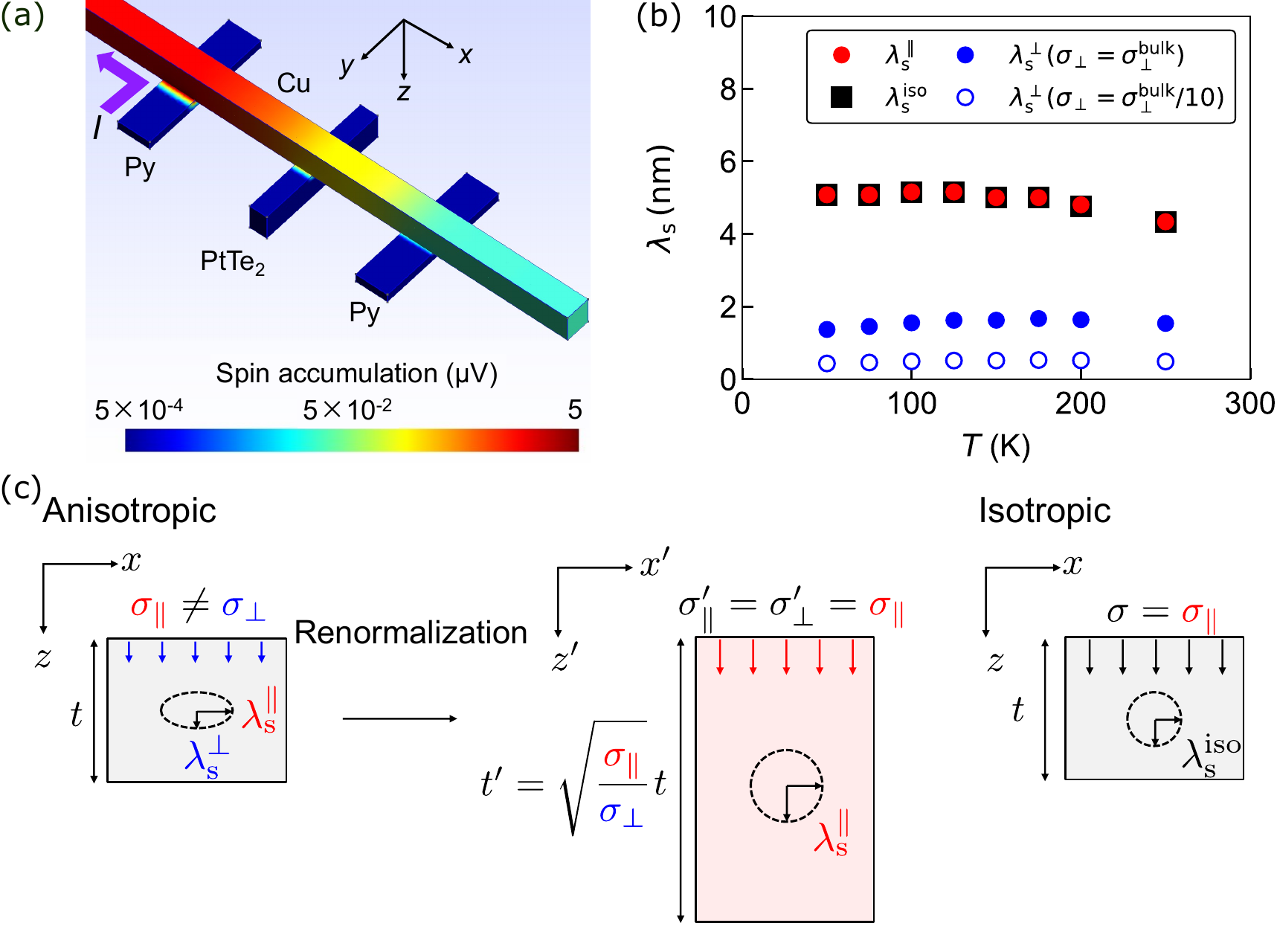}
    \caption{
          (a) Typical 3D model with renormalized $z$-axis for a PtTe$_2$ nanowire and the calculated spin accumulation voltage in the NLSV geometry.
          (b) Spin diffusion length of PtTe$_2$ as a function of temperature. 
          The black square shows the spin diffusion length $\lambda_{\mathrm{s}}^{\mathrm{iso}}$ obtained from the conventional isotropic analysis, assuming $\sigma = \sigma_{\parallel}$.
          The red circle indicates the in-plane spin diffusion length $\lambda_{\mathrm{s}}^{\parallel}$ obtained from the anisotropic analysis. 
          The blue closed (open) circles indicate the out-of-plane spin diffusion length $\lambda_{\mathrm{s}}^{\perp}$ obtained by assuming the out-of-plane conductivity $\sigma_{\perp}=\sigma_{\perp}^{\mathrm{bulk}}$ ($\sigma_{\perp}^{\mathrm{bulk}}/10$).
          (c) Comparison between $\lambda_{\mathrm{s}}^{\mathrm{iso}}$ and $\lambda_{\mathrm{s}}^{\parallel}$. 
          Through the renormalization in the anisotropic model, the conductivity is transformed to become isotropic for the in-plane and out-of-plane directions (i.e., $\sigma_{\parallel}' = \sigma_{\perp}' = \sigma_{\parallel}$) and the effective thickness is rescaled by a factor of $\sqrt{\sigma_{\parallel}/\sigma_{\perp}}$.
          This makes the thickness thicker compared to the isotropic case. 
          In the present case, the actual thickness ($t = 26$~nm in Sample A) is much larger than the spin diffusion length. Thus, this renormalization has little impact on the total spin absorption. 
        }
    \label{fig:5}
  \end{center}
\end{figure}

As the first step, we calculate the spin diffusion length in a NLSV device 
shown in Fig.~\ref{fig:4}(a) using a 3D model.
We used Gmsh, a 3D element mesh generator, and GetDP as a FEM solver~\cite{Geuzaine2009, Dular1999} for computing the nonlocal spin signal.
In order to renormalize the $z$-axis, we multiplied the thickness $t$ of PtTe$_2$ by the factor of $\sqrt{\sigma_{\parallel}/\sigma_{\perp}}$.
Based on Eqs.~\eqref{eq:2a} and \eqref{eq:2b}, we computed the nonlocal spin signals for both parallel and antiparallel magnetic configurations~\cite{Laczkowski2019}.
We determined the spin diffusion length $\lambda_{\mathrm{s}}^{\parallel}$ such that the experimentally observed ratio $\Delta R_{\mathrm{s}}^{\mathrm{with}}/\Delta R_{\mathrm{s}}^{\mathrm{without}}$ can be reproduced.
Subsequently, $\lambda_{\mathrm{s}}^{\perp}$ was obtained by transforming back to the original axis.
To investigate the effect of the anisotropy, we also computed the spin diffusion length assuming a conventional isotropic conductivity, i.e., $\sigma = \sigma_{\parallel}$.
Hereafter, we define the spin diffusion length obtained under this assumption as $\lambda_{\mathrm{s}}^{\mathrm{iso}}$.

In Fig.~\ref{fig:5}(a) we show spin accumulation calculated by the 3D model with the renormalized axis.
Figure~\ref{fig:5}(b) shows the temperature dependence of the spin diffusion length calculated with the anisotropic ($\lambda_{\mathrm{s}}^{\parallel}$ and $\lambda_{\mathrm{s}}^{\perp}$) and isotropic analyses $\lambda_{\mathrm{s}}^{\mathrm{iso}}$ for Sample A.
Notably, $\lambda_{\mathrm{s}}^{\parallel}$ and $\lambda_{\mathrm{s}}^{\mathrm{iso}}$ show nearly the same values.
This behavior can be interpreted by using the renormalized axis, 
as illustrated in Fig.~\ref{fig:5}(c).
Through the renormalization in the anisotropic model, the conductivity is transformed to become isotropic for the in-plane and out-of-plane directions (i.e., $\sigma_{\parallel}' = \sigma_{\perp}' = \sigma_{\parallel}$) and the effective thickness is rescaled by a factor of $\sqrt{\sigma_{\parallel}/\sigma_{\perp}}$, as shown in Fig.~\ref{fig:5}(c).
However, since the thickness of Sample A ($t = 26$~nm) is much thicker than the spin diffusion length, the total spin absorption remains almost unchanged even when the thickness is changed from $t$ to $\sqrt{\sigma_{\parallel}/\sigma_{\perp}}\,t$.
Consequently, the extracted spin diffusion lengths $\lambda_{\mathrm{s}}^{\mathrm{iso}}$ and $\lambda_{\mathrm{s}}^{\parallel}$ are nearly identical.
As mentioned in Sec.~\ref{sec:2}, the out-of-plane spin diffusion length $\lambda_{\mathrm{s}}^{\perp}$ is determined by $\mathrm{\lambda}_{\mathrm{s}}^{\perp}=\sqrt{\sigma_{\perp}/\sigma_{\parallel}}\,\mathrm{\lambda}_{\mathrm{s}}^{\parallel}$. 
Therefore, if we assume $\sigma_{\perp}=\sigma_{\perp}^{\mathrm{bulk}}/10$, 
$\lambda_{\mathrm{s}}^{\perp}$ is reduced by a factor of $\sqrt{10}$, compared to 
that obtained with $\sigma_{\perp}=\sigma_{\perp}^{\mathrm{bulk}}$.
According to a previous measurement using spin-torque ferromagnetic resonance~\cite{Wang2024}, the spin diffusion length of MBE-grown PtTe$_2$ was approximately 5~nm. Given that the spin current in this technique primarily flows along the out-of-plane direction, the reported value is likely to be overestimated.

All the above calculations were performed using a 3D simulation.  
However, we found that similar results can be obtained using a 1D model derived 
from the 1D spin diffusion equation.  
Although the 1D model is applicable only to the SHE materials characterized by a short spin diffusion length, it provides a useful approximation.
Details of the 1D approach are given in Appendix~\ref{appendix:H}.

\begin{figure}
  \begin{center}
    \includegraphics[width = \linewidth]{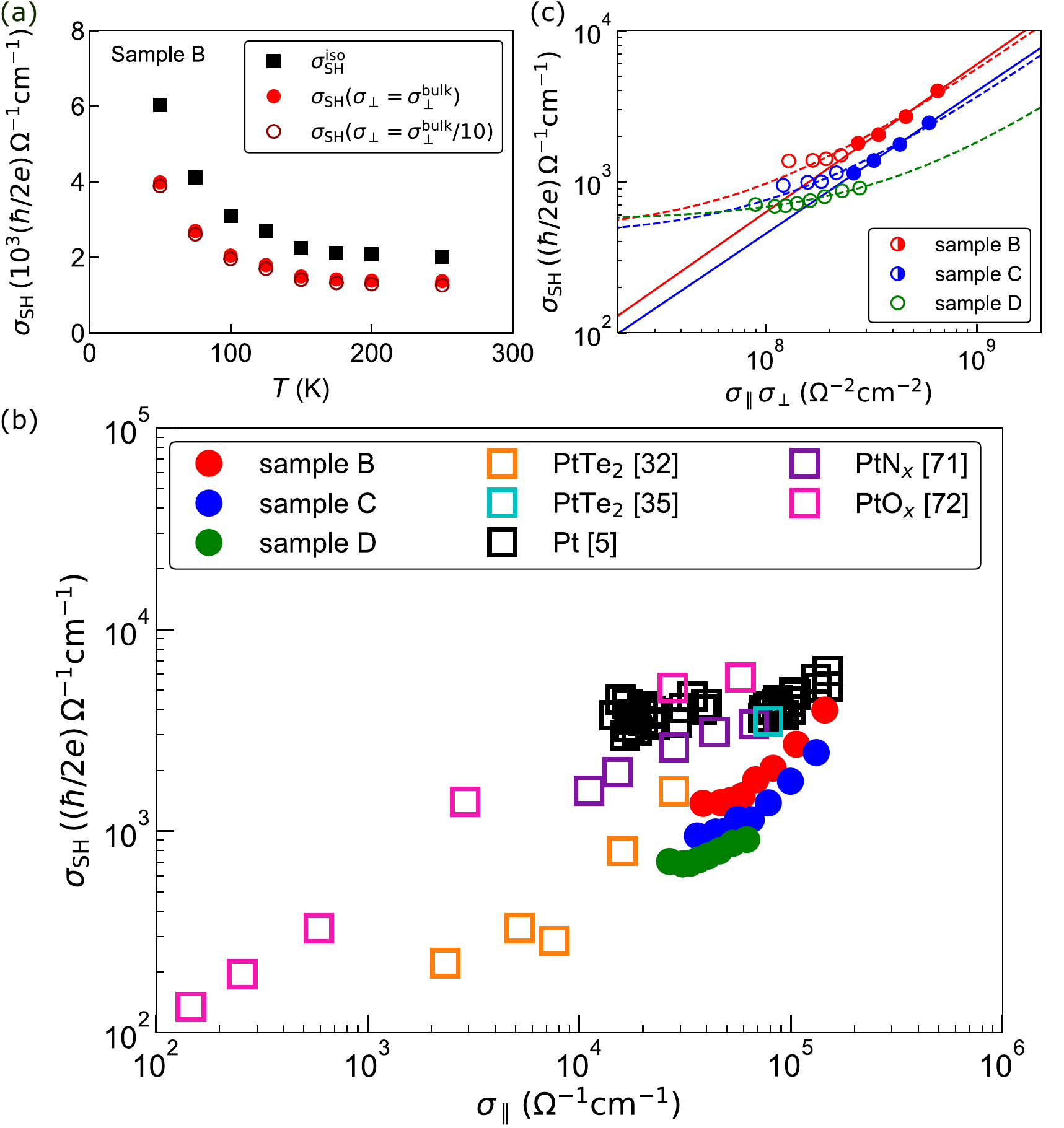}
    \caption{
        (a) Temperature dependence of the spin Hall conductivity of PtTe$_2$ for Sample B. The black squares represent calculated results from the isotropic analysis.
        The red (brown) circles represent calculated results from the anisotropic analysis by assuming the out-of-plane conductivity $\sigma_{\perp}=\sigma_{\perp}^{\mathrm{bulk}}$ ($\sigma_{\perp}^{\mathrm{bulk}}/10$).
        (b) Log-log plot of the spin Hall conductivity as a function of $\sigma_{\parallel}$. Literature data for PtTe$_2$~\cite{Xu2020, Wang2024} and other Pt-based materials (Pt, PtN$_x$, PtO$_x$)~\cite{Sagasta2016, Soya2021, Moriya2022} are also included for comparison.
        (c) Log-log plot of the spin Hall conductivity as a function of $\sigma_{\parallel}\sigma_{\perp}$ for Samples B (red), C (blue), and D (green). 
        Closed circles represent a high-conductivity regime and are fitted using $\sigma_{\mathrm{SH}} \propto (\sigma_{\parallel}\sigma_{\perp})^{\gamma}$.
        Open circles represent the remaining data points outside this regime. Dashed lines represent a fit to the entire dataset (both closed and open circles) using
        $\sigma_{\mathrm{SH}} = \sigma_{\mathrm{SH}}^{\mathrm{int}} + a'\rho_{\parallel 0} \sigma_{\parallel} \sigma_{\perp}$.
        }
    \label{fig:6}
  \end{center}
\end{figure}

In the next step, we analyze the ISHE.
The spin Hall angle $\theta_{\mathrm{SH}} = \sigma_{\mathrm{SH}} / \sigma$ is often used to characterize the spin-charge conversion.  
Although it serves as a useful indicator of spin-charge conversion efficiency in application,
it does not reflect a fundamental physical property, especially in systems with an anisotropic conductivity,  
where $\sigma_{\mathrm{SH}}/\sigma_{\parallel} \neq \sigma_{\mathrm{SH}}/\sigma_{\perp}$.  
Therefore, we use the spin Hall conductivity $\sigma_{\mathrm{SH}}$ itself, which is a more intrinsic quantity, to evaluate the spin-charge conversion.
As we have done for the spin diffusion length, we consider 
two cases to investigate the effect of the anisotropy: $\sigma_{\mathrm{SH}}$ in the anisotropic system, and $\sigma_{\mathrm{SH}}^{\mathrm{iso}}$ where the isotropic conductivity $\sigma = \sigma_{\parallel}$ is assumed.
Figure~\ref{fig:6}(a) shows the temperature dependence of the spin Hall conductivities obtained from the anisotropic ($\sigma_{\mathrm{SH}}$) and isotropic ($\sigma_{\mathrm{SH}}^{\mathrm{iso}}$) analyses.
The difference between the two $\sigma_{\mathrm{SH}}$ values obtained with $\sigma_{\perp}=\sigma_{\perp}^{\mathrm{bulk}}$ and $\sigma_{\perp}=\sigma_{\perp}^{\mathrm{bulk}}/10$ is small.
The results satisfy the relation  
$\sigma_{\mathrm{SH}} < \sigma_{\mathrm{SH}}^{\mathrm{iso}}$,
which can be attributed to the suppression of the shunting effect by the Cu electrode in the presence of conductivity anisotropy.
Such behavior is not unique to PtTe$_2$, but is expected to occur more generally in other anisotropic layered materials.
Therefore, in previous studies where the conductivity anisotropy was neglected by assuming $\sigma = \sigma_{\parallel}$, the spin Hall conductivity might have been overestimated.

Figure~\ref{fig:6}(b) shows a log-log plot of $\sigma_{\mathrm{SH}}$ as a function of $\sigma_{\parallel}$, where we also include some data of PtTe$_2$ from the literature~\cite{Xu2020, Wang2024} and other Pt-based materials (Pt, PtN$_x$, PtO$_x$)~\cite{Sagasta2016, Soya2021, Moriya2022} for comparison.
The reason for using $\sigma_{\parallel}$ as the horizontal axis is that the charge current flows in the in-plane direction, and most of the scattering events should depend on the in-plane conductivity.
Our results exhibit the same trend as those from previous studies on PtTe$_2$~\cite{Xu2020}, although the values presented in our work are slightly smaller. 
Moreover, compared to PtTe$_2$ films fabricated by CVD and MBE~\cite{Xu2020, Wang2024}, the PtTe$_2$ flakes used in this study exhibit higher conductivities due to the high crystal quality.

We now focus on the origin of the SHE in PtTe$_2$.
The SHE consists of intrinsic and extrinsic mechanisms, and the total $\sigma_{\mathrm{SH}}$ can be expressed as:
$\sigma_{\mathrm{SH}}=\sigma_{\mathrm{SH}}^{\mathrm{int}}+\sigma_{\mathrm{SH}}^{\mathrm{ext}}$~\cite{Nagaosa2010}.
The intrinsic mechanism originates from the Berry curvature of the electronic band structure and is relatively insensitive to impurity scattering.
As a result, when the intrinsic mechanism is dominant, $\sigma_{\mathrm{SH}}$ remains nearly constant with respect to the conductivity of PtTe$_2$.
According to Fig.~\ref{fig:6}(b), $\sigma_{\mathrm{SH}}$ remains nearly constant in the low-conductivity regime, indicating a dominant intrinsic contribution.  
By contrast, in the high-conductivity regime, $\sigma_{\mathrm{SH}}$ increases with increasing the conductivity, suggesting a crossover from the intrinsic to 
extrinsic mechanisms.  
Such behavior is consistent with a previously reported SHE in Pt, where 
a crossover from the moderately dirty regime to the superclean regime was demonstrated by tuning the conductivity~\cite{Sagasta2016}.

To gain deeper insight into this behavior, we consider the extrinsic contributions arising from scattering sources such as impurities and phonons.
There are two extrinsic mechanisms: side-jump and skew scattering.
The microscopic theory of the side-jump mechanism gives a constant spin Hall conductivity with respect to the conductivity. 
Using the inverse matrix relation $-\rho_{\mathrm{SH}}=\sigma_{\mathrm{SH}}/(\sigma_{\parallel}\sigma_{\perp}+\sigma_{\mathrm{SH}}^2)
\approx\sigma_{\mathrm{SH}}/\sigma_{\parallel}\sigma_{\perp}$, which is valid when $\sigma_{\mathrm{SH}}^2 \ll \sigma_{\parallel} \sigma_{\perp}$ [see Appendix~\ref{appendix:B}],
the spin Hall resistivity due to the side-jump is proportional to $\rho_{\parallel}\rho_{\perp}$. 
In contrast, the spin Hall resistivity due to skew scattering should be 
proportional to longitudinal resistivity $\rho_{\parallel}$~\cite{Nagaosa2010}.
Based on the scaling theory developed for the anomalous Hall effect~\cite{Tian2009}, the impurity contribution is more dominant than the phonon contribution.
Therefore, these extrinsic contributions can be described 
using Matthiessen’s rule as follows:
$-\rho_{\mathrm{SH}}^{\mathrm{ext}}=a'\rho_{\parallel0}+\beta\rho_{\parallel 0}\rho_{\perp 0}$,
where $\rho_{\parallel 0}$ and $\rho_{\perp 0}$ are the residual components of the in-plane resistivity and out-of-plane resistivities, respectively.
We note that the side-jump contribution is neglected ($\beta \approx 0$), since it is often smaller than the skew scattering in high-conductivity metals~\cite{Isasa2015, Sagasta2016, Karnad2018}.
Thus, the total spin Hall conductivity is expressed as
$\sigma_{\mathrm{SH}}=\sigma_{\mathrm{SH}}^{\mathrm{int}}+a'\rho_{\parallel0}\sigma_{\parallel}\sigma_{\perp}$.

We plotted $\sigma_{\mathrm{SH}}$ vs $\sigma_{\parallel}\sigma_{\perp}$ on a double logarithmic scale, as shown in Fig.~\ref{fig:6}(c).
To examine the scaling behavior in the extrinsic regime, the data points between 50~K and 125~K for Samples B and C (closed circles) were fitted using the relation $\sigma_{\mathrm{SH}} \propto (\sigma_{\parallel}\sigma_{\perp})^{\gamma}$. 
The fitted curves are shown as solid lines.
We obtained exponents of $\gamma = 0.98\pm 0.07$ for Sample B and $\gamma = 0.95\pm 0.03$ for Sample C.
Since both results show exponents $\gamma \approx 1$, this indicates that the $\sigma_{\mathrm{SH}}$ in the higher conductivity region is dominated by the impurity skew scattering.
We then applied the expression
$\sigma_{\mathrm{SH}} = \sigma_{\mathrm{SH}}^{\mathrm{int}} + a' \rho_{\parallel 0} \sigma_{\parallel} \sigma_{\perp}$,
to fit the full dataset, including Sample D and the low-conductivity regions of Samples B and C (open circles), as shown by dashed lines.
These data points are likely located in a mixed regime where both extrinsic and intrinsic contributions are present, making precise analysis challenging. 
Moreover, no universal behavior was found for the coefficient $a'$ in this study [see Appendix~\ref{appendix:D} for details].
Nevertheless, the fitting yields an average value of $\sigma_{\mathrm{SH}}^{\mathrm{int}}=(4.8 \pm 0.5)\times 10^2(\hbar/2e)\Omega^{-1}\mathrm{cm}^{-1}$.

To verify the reliability of the obtained $\sigma_{\mathrm{SH}}^{\mathrm{int}}$, we performed first-principles calculations of the spin Hall conductivity $\sigma_{\mathrm{SH}}$
[see Appendix~\ref{appendix:I} for details of computations].
As a result, we obtained
$\sigma^{y}_{zx} = 1.8 \times 10^2\,(\hbar/2e)\mathrm{\Omega^{-1}cm^{-1}}$ ,
which is three times smaller than our experimental value and nine times smaller than our calculation for Pt ($4.4 \times 10^3\,(\hbar/2e)\mathrm{\Omega^{-1}cm^{-1}}$).
In the following, we discuss possible reasons why the experimentally obtained $\sigma_{\mathrm{SH}}^{\mathrm{int}}$ differs by a factor of three from the first-principles calculation.
One possible reason is 
the interfacial spin-charge conversion, such as the Rashba-Edelstein effect (REE).
Indeed, local-Rashba splitting has been experimentally observed in PtTe$_2$~\cite{Deng2019, Cai2024, Feng2025}. 
Thus, the interfacial conversion efficiency could be enhanced 
by the REE, in addition to the bulk SHE of PtTe$_{2}$.
This is supported by theoretical predictions that monolayer PtTe$_2$ exhibits a high spin Hall conductivity of $2.5 \times 10^3\,(\hbar/2e)\mathrm{\Omega^{-1}cm^{-1}}$ due to Rashba splitting~\cite{Shao2023}.
Furthermore, according to recent studies on large-area PtTe$_2$ films~\cite{Xu2020, Yadav2024}, the thickness dependence of spin-charge conversion efficiency in PtTe$_2$ is distinct from that observed in Pt, suggesting the dominance of the interfacial spin-charge conversion.
This scenario is consistent with the extremely short out-of-plane spin diffusion length ($\lambda_{\mathrm{s}}^{\perp} =$ 1-2~nm) obtained in our measurement, confirming that the spin-charge conversion is spatially confined to the interface.
Therefore, the experimentally obtained spin Hall conductivity should be interpreted as an effective value resulting from the combination of the bulk SHE and REE. 
The comparison with the first-principles calculations highlights the significant role of interfacial effects in enhancing the total conversion efficiency in PtTe$_2$. 
Further experiments, such as systematic thickness-dependent measurements, are needed to separate the bulk SHE from the interfacial REE in future work. 

\section{\label{section6}Conclusion}
We have quantitatively evaluated the anisotropic spin transport properties of PtTe$_2$ by combining 3D FEM with NLSV devices.
Our theoretical model enables accurate extraction of spin diffusion lengths along both in-plane and out-of-plane directions.
We reveal the following relations 
$\lambda_{\mathrm{s}}^{\perp} < \lambda_{\mathrm{s}}^{\mathrm{iso}} \leq \lambda_{\mathrm{s}}^{\parallel}$ and
$\sigma_{\mathrm{SH}} < \sigma_{\mathrm{SH}}^{\mathrm{iso}}$, indicating that the conventional isotropic model ($\sigma=\sigma_{\parallel}$) leads to overestimations of the out-of-plane spin diffusion length and spin Hall conductivity.
Moreover, our results revealed a crossover from intrinsic to extrinsic SHE regimes as the conductivity increases.
These findings provide crucial insight into spin-charge conversion mechanisms in layered materials and highlight the importance of considering anisotropic spin diffusions in the design of future spintronic devices.

\section{Acknowledgments}
We gratefully acknowledge A. Tsukazaki and J. Matsuno for the insightful suggestions.
We thank F. Casanova, A. Marty, and F. Bonell for their fruitful discussions.
This work was supported by Japan Society for the Promotion of Science KAKENHI (Grants Nos.~JP23H00257, JP23KJ1502, JP23K03275, JP25H00841), JST PRESTO (Grant No.~JPMJPR2452) and JST FOREST (Grant No.~JPMJFR2134).

\appendix

\section{Anisotropy of spin relaxation time $\tau_{\mathrm{sf}}$\label{appendix:A}}
There are two distinct types of anisotropy that can be discussed in the context of spin relaxation.
One is spin-orientation anisotropy, which refers to the case where the spin relaxation time depends on the spin orientation, i.e., $\tau_{\mathrm{sf}}$ for spins polarized along the $x$-direction differs from that along the $z$-direction. 
This type of anisotropy has been extensively studied in layered materials such as graphene and black phosphorus~\cite{Raes2016, Leutenantsmeyer2018, Xu2018, Cording2024, Yalcin2024}.
The other is spatial anisotropy, which refers to the case where the spin relaxation time depends on the direction in which spins propagate, even for a fixed spin orientation.

In the present case, the latter anisotropy is dominant and does not affect $\tau_{\mathrm{sf}}$ as an independent parameter. 
Intuitively, this is because the spin relaxation time $\tau_{\mathrm{sf}}$ characterizes how quickly a spin loses its orientation due to spin-flip scattering events -- it is a local property of the electron at a given position, independent of the direction in which the electron is traveling. 
On the other hand, the diffusion coefficient $D$ describes how fast electrons propagate through the material, which naturally depends on the crystallographic direction in an anisotropic system. 
Therefore, $D$ can be anisotropic, while $\tau_{\mathrm{sf}}$ remains isotropic with respect to the spatial direction of propagation.

\section{Spin diffusion equations in anisotropic systems for SHE and ISHE\label{appendix:B}}
The SHE is described by the linear response of the spin current $J_{j}^i$ to the applied electric field $E_k$, given by
$J_{j}^i = \sigma_{jk}^{i,\mathrm{cs}} E_k$.
Here $i$ denotes the spin polarization axis, $j$ denotes the spin current direction, $k$ denotes the direction of the applied electric field, and $\sigma_{jk}^{i,\mathrm{cs}}$ is the spin Hall conductivity tensor that characterizes charge-to-spin conversion.
On the other hand, the ISHE is expressed through the gradient of the spin chemical potential $E_{k}^i\,(\equiv -(1/e)\partial_k\mu_{s}^i)$ and charge current $J_{j}$, as $J_{j} = \sigma_{jk}^{i, \mathrm{sc}} E_{k}^i$.
Here, $\sigma_{jk}^{i, \mathrm{sc}}$ is the inverse spin Hall conductivity tensor that characterizes spin-to-charge conversion. 

In conventional nonlocal spin valve devices, spin polarization is set along the $x$ axis, and $\sigma_{yz}^x$ and $\sigma_{zy}^x$ are studied as the SHE and ISHE in the $yz$ plane, respectively.
Based on the symmetry of PtTe$_2$, the in-plane electrical conductivity is isotropic ($\sigma_{xx} = \sigma_{yy}$).
Accordingly, we define the in-plane and out-of-plane conductivities as $\sigma_{\parallel} (\equiv \sigma_{xx} = \sigma_{yy})$ and $\sigma_{\perp} (\equiv \sigma_{zz})$, respectively.
Consequently, the SHE and ISHE in the $yz$ plane can be expressed as

\begin{align}
    \begin{pmatrix} 
    J_y \\ 
    J_z^x 
    \end{pmatrix}
    &=
    \begin{pmatrix} 
    \sigma_{\parallel} & \sigma_{yz}^{x,\mathrm{sc}} \\ 
    \sigma_{zy}^{x,\mathrm{cs}} & \sigma_{\perp} 
    \end{pmatrix}
    \begin{pmatrix} 
    E_y \\ 
    E_z^x 
    \end{pmatrix},\\
    \begin{pmatrix} 
    J_z \\ 
    J_y^x 
    \end{pmatrix}
    &=
    \begin{pmatrix} 
    \sigma_{\perp} & \sigma_{zy}^{x,\mathrm{sc}} \\ 
    \sigma_{yz}^{x,\mathrm{cs}} & \sigma_{\parallel} 
    \end{pmatrix}
    \begin{pmatrix} 
    E_z \\ 
    E_y^x 
    \end{pmatrix}.
\end{align}
It is known that the $\sigma_{yz}^{x,\mathrm{cs}} = -\sigma_{zy}^{x,\mathrm{cs}}=\sigma_{yz}^{x,\mathrm{sc}} = -\sigma_{zy}^{x,\mathrm{sc}}$ holds when the system has a rotational symmetry around the $x$-axis, such as threefold or fourfold rotational symmetries. 
However, in many layered materials, this symmetry is broken due to the anisotropy between the $y$ and $z$ directions. 
In the case of PtTe$_2$ (space group $P\bar{3}m1$, No.~164), the symmetry constraints imposed by threefold rotoinversion and mirror operations result in the following nonzero spin Hall conductivity tensors~\cite{Roy2022}:
\begin{align}
    \sigma^{x,\mathrm{cs}}&=
    \begin{pmatrix}
        0 & \sigma_{xy}^{x,\mathrm{cs}} & 0 \\
        \sigma_{yx}^{x,\mathrm{cs}} & 0 & \sigma_{yz}^{x,\mathrm{cs}} \\
        0 & \sigma_{zy}^{x,\mathrm{cs}} & 0
    \end{pmatrix}
    ,\notag\\
    \sigma^{y,\mathrm{cs}}&=
    \begin{pmatrix}
        \sigma_{xx}^{y,\mathrm{cs}} & 0 & \sigma_{xz}^{y,\mathrm{cs}} \\
        0 & \sigma_{yy}^{y,\mathrm{cs}} & 0 \\
        \sigma_{zx}^{y,\mathrm{cs}}  &0 & 0
    \end{pmatrix}
    ,\notag\\
    \sigma^{z,\mathrm{cs}}&=
    \begin{pmatrix}
        0 &\sigma_{xy}^{z,\mathrm{cs}} &0 \\
        \sigma_{yx}^{z,\mathrm{cs}} &0 &0 \\
        0 &0 &0
    \end{pmatrix}.
\end{align}
Note that in this study, the $a$-axis of PtTe$_2$ is oriented along the $y$-axis in the NLSV, corresponding to $90^\circ$ rotation relative to Ref.~\cite{Roy2022}.
Additionally, these tensor components satisfy 
$\sigma_{xx}^{y,\mathrm{cs}}=\sigma_{xy}^{x,\mathrm{cs}}=\sigma_{yx}^{x,\mathrm{cs}}=-\sigma_{yy}^{y,\mathrm{cs}},\,\sigma_{yz}^{x,\mathrm{cs}}=-\sigma_{xz}^{y,\mathrm{cs}},\, \sigma_{zy}^{x,\mathrm{cs}}=-\sigma_{zx}^{y,\mathrm{cs}},\,\sigma_{xy}^{z,\mathrm{cs}}=-\sigma_{yx}^{z,\mathrm{cs}}$, leaving only four independent spin Hall conductivity tensors. 
Therefore, in PtTe$_2$, the SHE and ISHE tensors are not strictly antisymmetric, i.e.,
$\sigma_{yz}^{x,\mathrm{cs}} \neq -\sigma_{zy}^{x,\mathrm{cs}}$ and
$\sigma_{yz}^{x,\mathrm{sc}} \neq -\sigma_{zy}^{x,\mathrm{sc}}$.
However, 
we obtained $\sigma_{yz}^{x,\mathrm{cs}}=1.85\times10^2\,(\hbar/2e)\, \Omega^{-1}\mathrm{cm}^{-1}$ and $\sigma_{zx}^{y,\mathrm{cs}}=1.81\times10^2\,(\hbar/2e)\, \Omega^{-1}\mathrm{cm}^{-1}$ from our first-principles calculations (see Appendix~\ref{appendix:I} for details of computations).
By applying the symmetry relation $\sigma_{zy}^{x,\mathrm{cs}}=-\sigma_{zx}^{y,\mathrm{cs}}$ to these values, we find that $\sigma_{yz}^{x,\mathrm{cs}}\approx-\sigma_{zy}^{x,\mathrm{cs}}$. The difference in magnitude is only 2\%, confirming that the SHE tensor is nearly antisymmetric.
Therefore, provided that the Onsager reciprocity relation $\sigma_{ij}^{k,\mathrm{cs}}=-\sigma_{ji}^{k,\mathrm{sc}}$ holds, $\sigma_{\mathrm{SH}}\equiv\sigma_{zy}^{x,\mathrm{cs}} = -\sigma_{yz}^{x,\mathrm{cs}} = \sigma_{zy}^{x,\mathrm{sc}} = -\sigma_{yz}^{x,\mathrm{sc}}$.
With this approximation,
\begin{align}
    \begin{pmatrix}
    J_y \\
    J_z^x
    \end{pmatrix}
    =
    \begin{pmatrix}
    \sigma_{\parallel} & -\sigma_{\mathrm{SH}} \\
    \sigma_{\mathrm{SH}} & \sigma_{\perp}
    \end{pmatrix}
    \begin{pmatrix}
    E_y \\
    E_z^x
    \end{pmatrix},~\label{eq:ohm_sigma1} \\
    \begin{pmatrix}
    J_z \\
    J_y^x
    \end{pmatrix}
    =
    \begin{pmatrix}
    \sigma_{\perp} & \sigma_{\mathrm{SH}} \\
    -\sigma_{\mathrm{SH}} & \sigma_{\parallel}
    \end{pmatrix}
    \begin{pmatrix}
    E_z \\
    E_y^x
    \end{pmatrix}.~\label{eq:ohm_sigma2}
\end{align}

By adding these SHE and ISHE terms to the following spin diffusion equations, 
\begin{equation}
\left\{
\begin{aligned}
   \nabla\cdot \left[\sigma\nabla(\mu_{\uparrow}+\mu_{\downarrow})\right] &= 0 \\
   \frac{1}{2e}\nabla\cdot 
      \left[\sigma\nabla(\mu_{\uparrow}-\mu_{\downarrow})\right]
      &= \frac{e\rho(\epsilon_{\mathrm{F}})}{\tau_{\mathrm{sf}}}(\mu_{\uparrow}-\mu_{\downarrow})
\end{aligned}
\right.
\end{equation}
we obtain Eqs.~\eqref{eq:1a} and~\eqref{eq:1b}.
From Eq.~\eqref{eq:ohm_sigma1} or \eqref{eq:ohm_sigma2} and the relation \(\rho = \sigma^{-1}\), we also obtain
\begin{align}
   \begin{pmatrix}
        \rho_{\parallel} & \rho_{\mathrm{SH}} \\
        -\rho_{\mathrm{SH}} & \rho_{\perp} \\
   \end{pmatrix}
   =
   \frac{1}{\sigma_{\perp}\sigma_{\parallel}+\sigma_{\mathrm{SH}}^2}
   \begin{pmatrix}
        \sigma_{\perp} & -\sigma_{\mathrm{SH}} \\
        \sigma_{\mathrm{SH}} & \sigma_{\parallel} \\
   \end{pmatrix}.\label{eq:ohm_rho}
\end{align}

\section{Estimation of interfacial resistance $R_{\mathrm{int}}$\label{appendix:C}}
\begin{figure}[htb]
  \begin{center}
    \includegraphics[width = \linewidth]{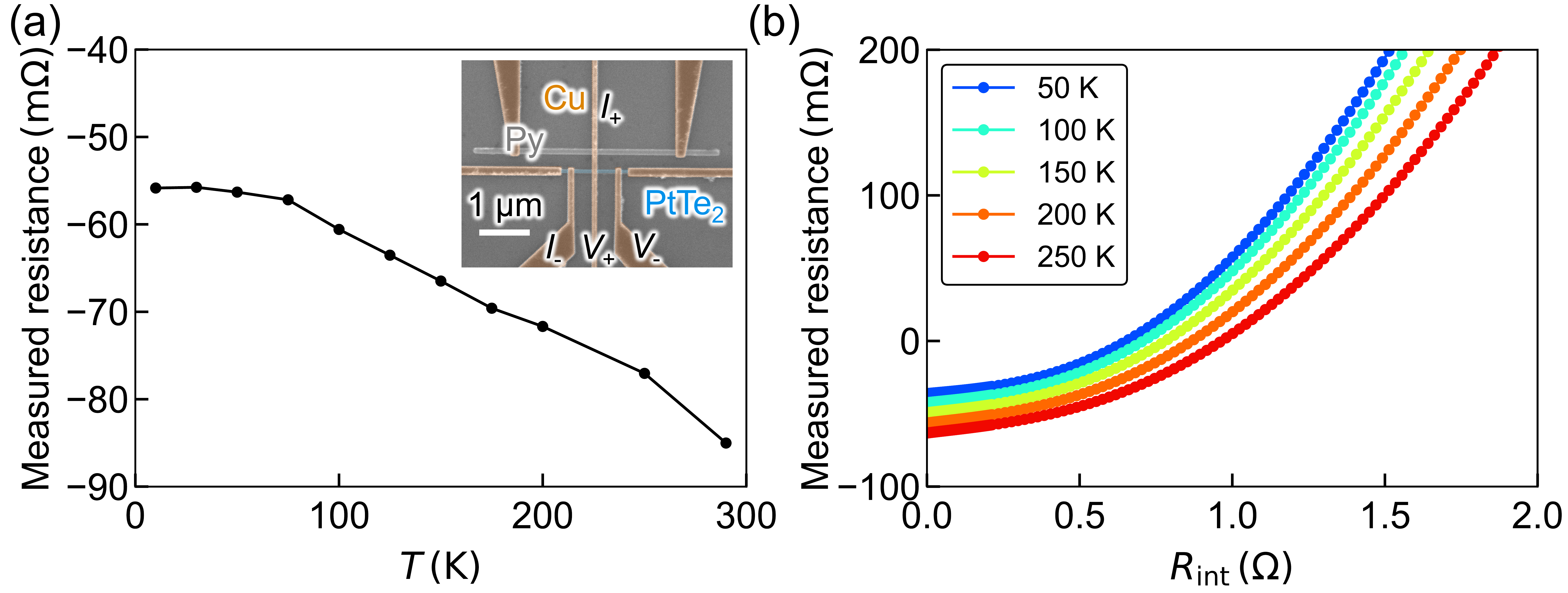}
    \caption{
      (a) Temperature dependence of the resistance measured in the configuration used to extract $R_{\mathrm{int}}$ shown in the inset. 
      (b) 3D-FEM simulations of the measured resistance as a function of $R_{\mathrm{int}}$ at various temperatures.
    }
    \label{fig:C}
  \end{center}
\end{figure}

To estimate the interfacial resistance $R_{\mathrm{int}}$ at the Cu/PtTe$_2$ interface, we used the measurement setup shown in the inset of Fig.~\ref{fig:C}(a). 
The temperature dependence of the resistance measured in this configuration is presented in Fig.~\ref{fig:C}(a). 
The resistance is negative over the entire temperature range, which is a characteristic feature of highly transparent interfaces. 
To confirm this, we performed a 3D-FEM simulation of this geometry. 
As shown in Fig.~\ref{fig:C}(b), the simulation indicates that a very low $R_{\mathrm{int}}$ value results in a small negative resistance, which is attributed to the current distribution across the interface. 
This result justifies the cleanliness of the interface.

\section{Device parameters and fitting results\label{appendix:D}}

\renewcommand{\arraystretch}{1.3}
\begin{table*}
\centering
\caption{Summary of geometric parameters and fitting results for all the devices.}
\label{tab:device_summary}
\begin{tabular}{ccccccc}
\hline
\hline
Sample &$t_{\mathrm{M}}$~(nm) & $w_{\mathrm{M}}$~(nm) & $d$~(nm) & $\rho_{\parallel0}$~$(\mathrm{\mu\Omega\cdot cm})$ & $a'$~$(\%)$ & $\sigma_{\mathrm{SH}}^{\mathrm{int}}\,\left((\hbar/2e)\Omega^{-1}\mathrm{cm}^{-1}\right)$ \\
\hline
A & 26 & 64 & 470 & -- & -- & -- \\
B & 27 & 140 & 540 & 5.3 & 48$\pm$3 & 460$\pm$130 \\
C & 36 & 101 & 500 & 6.1 & 26$\pm$2 & 430$\pm$80 \\
D & 20 & 75 & 410 & 14.5 & 4.4$\pm$0.4 & 550$\pm$20 \\
\hline
\hline
\end{tabular}
\end{table*}

This appendix summarizes the geometric parameters and fitting results for all the investigated devices, as presented in Table~\ref{tab:device_summary}. 
The table lists the middle wire thickness $t_{\mathrm{M}}$, width $w_{\mathrm{M}}$, and the gap distance $d$ between the injector ferromagnetic wire and the middle wire. 
It also includes the in-plane residual resistivity $\rho_{\parallel0}$ measured at $T = 10$~K. 
The fitting parameters $a'$ and $\sigma_{\mathrm{SH}}^{\mathrm{int}}$ correspond to the scaling factor for the resistivity dependence and the intrinsic spin Hall conductivity, respectively.
It should be noted that the relatively small value of $a'$ for Sample D is likely due to the limited number of data points in the high-conductivity region.

\section{Size effect of in-plane resistivity $\rho_{\parallel}$\label{appendix:E}}

\begin{figure}[htb]
  \begin{center}
    \includegraphics[width = \linewidth]{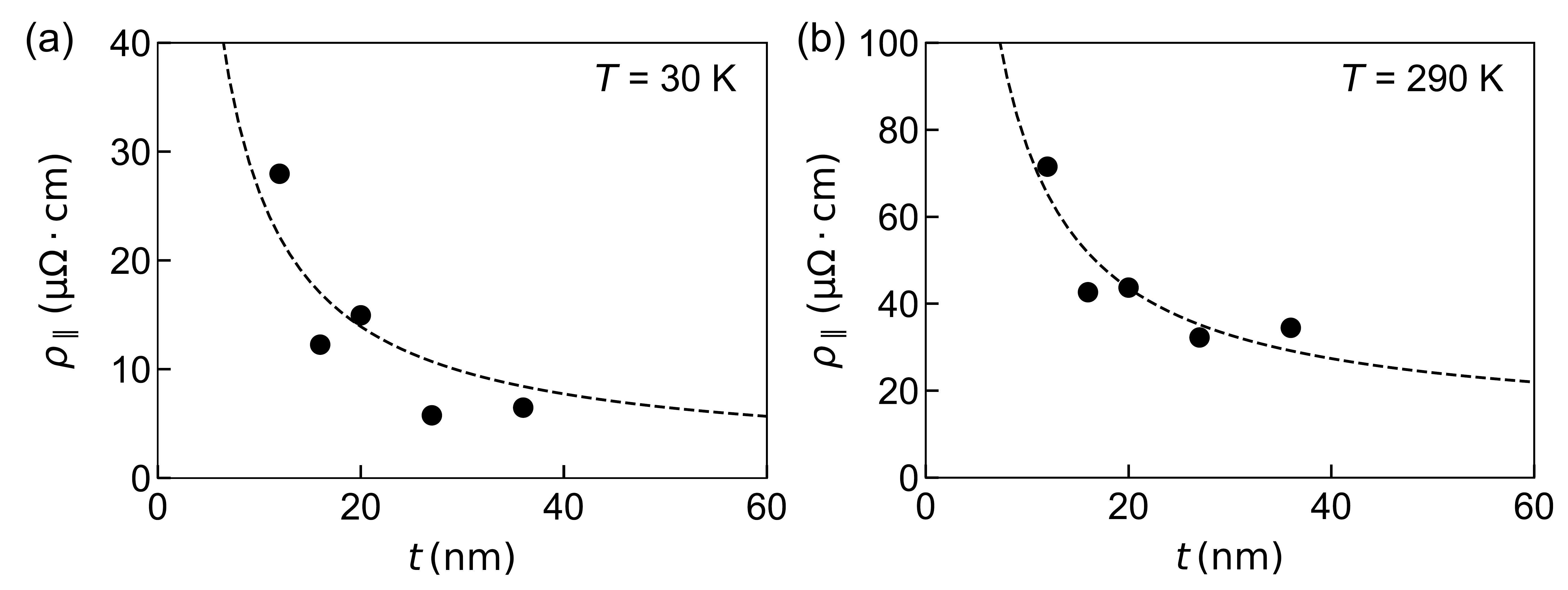}
    \caption{
        In-plane resistivity $\rho_{\parallel}$ as a function of thickness at (a) $T=30$~K and (b) 290~K. The dashed lines represent the fitting results.
        }
    \label{fig:E}
  \end{center}
\end{figure}

We have measured the size effect of the in-plane resistivity $\rho_{\parallel}$. 
As shown in Fig.~\ref{fig:E}, the thickness dependence of PtTe$_2$ is roughly described by the Fuchs-Sondheimer (FS) model~\cite{Sondheimer2001}, which is expressed as follows:
\begin{gather}
    \rho_{\parallel}=\left(1+\frac{3}{8}\frac{l}{t}\right)\rho_{\parallel}^{\mathrm{bulk}}.\label{eq:FS}
\end{gather}
We fitted the in-plane resistivity $\rho_{\parallel}$ of our PtTe$_2$ thin films using Eq.~\eqref{eq:FS}, with the elastic mean free path $l$ as a free parameter. The fitting yields $l=43$~nm at 30~K and $l=15$~nm at 290~K. 
These values are roughly consistent with the mean free path estimated from the carrier density reported in a previous study~\cite{Pavlosiuk2018}, which yielded values of $l=27$~nm at 10~K and $l=6$~nm at 300~K.
This result shows that due to the size effect, the in-plane resistivity of our 30~nm film is approximately seven times higher than the bulk value at low temperatures and three times higher at room temperature.

\section{Vertical contact method\label{appendix:F}}
\begin{figure}
  \begin{center}
    \includegraphics[width = 0.9\linewidth]{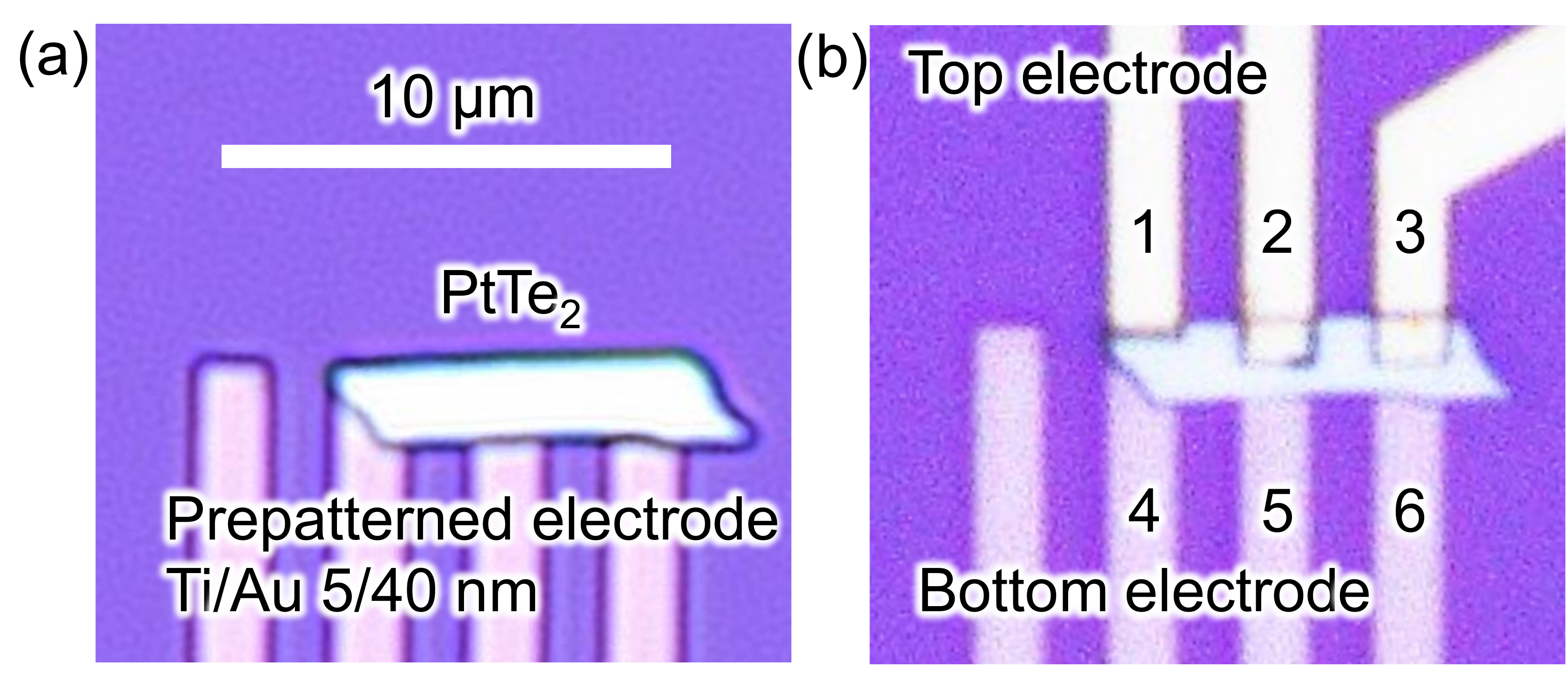}
    \caption{
        Optical microscope images showing the fabrication process of the vertical contact device.
        (a) After dry-transferring a PtTe$_2$ flake onto pre-patterned Ti/Au (5/40~nm) electrodes.
        (b) After depositing the top Ti/Au (5/100~nm) electrodes.
        }
    \label{fig:F}
  \end{center}
\end{figure}

To measure the out-of-plane resistivity, we adopted the vertical contact method.
A PtTe$_2$ film was dry-transferred onto pre-patterned Ti/Au (5/40~nm) electrodes, as shown in Fig.~\ref{fig:F}(a).
Subsequently, Ti/Au (5/100~nm) electrodes were deposited on the PtTe$_2$ film, as shown in Fig.~\ref{fig:F}(b). 
We measured the resistance in a four-terminal setup using two different contact schemes, with the notation ($I_+$–$I_-$–$V_+$–$V_-$) indicating the terminal connections.
For the in-plane configuration (1-2-4-5), the in-plane resistance was 8.60~$\mathrm{\Omega}$. 
For the out-of-plane configuration (1-4-2-5), the resistance was -0.75~$\mathrm{m\Omega}$.
This measurement method assumes that the out-of-plane resistance $ R_{\perp} $ is much higher than the in-plane resistance $ R_{\parallel} $, which implies that the in-plane direction is approximately equipotential and that no significant voltage drop occurs within the in-plane. 
This assumption can be expressed as
$\rho_{\perp} t/w \ell \gg \rho_{\parallel} \ell/w t$,
where $ \ell $ is the lateral size (3~$\mu$m in this device), $t$ is the thickness of PtTe$_2$ (40~nm in this device), and $w$ is the width.
However, since $\ell$ is much larger than $t$, this equipotential approximation may not be valid. 
As a result, $ R_{\parallel} $ is comparable to $ R_{\perp} $, and reliable resistance values can not be obtained.

\section{Out-of-plane resistivity $\rho_{\perp}$ measurement with bulk sample\label{appendix:G}}

\begin{figure}[htb]
  \begin{center}
    \includegraphics[width = \linewidth]{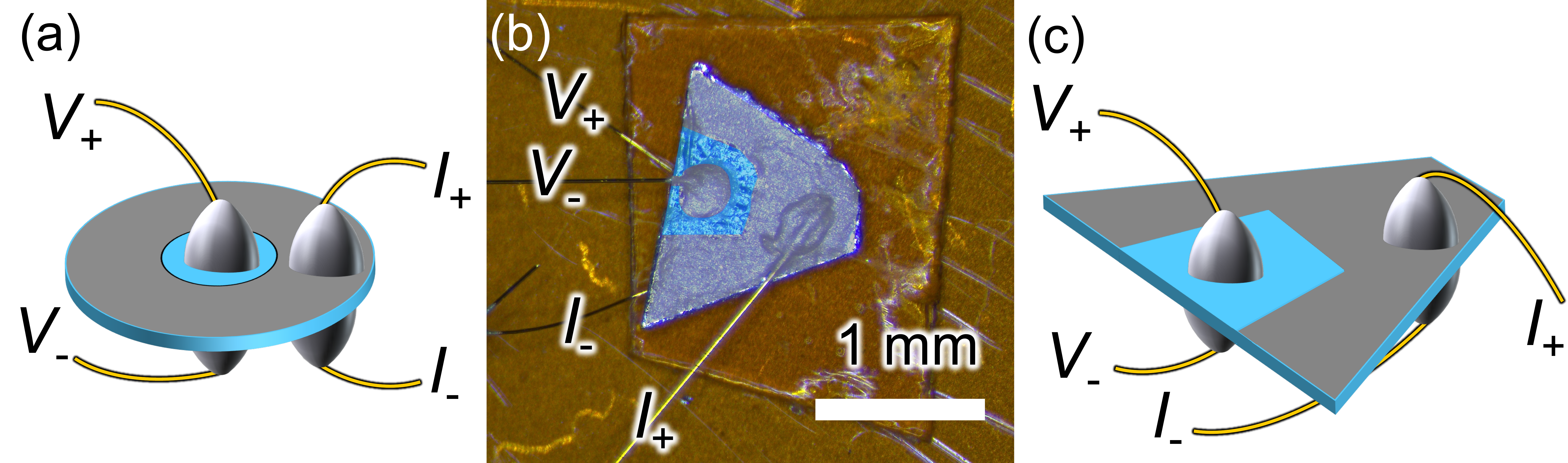}
    \caption{
      (a) Schematic image of the ideal electrode configuration (Corbino-geometry). The light-blue area represents PtTe$_2$, and the gray area represents silver pastes. 
      (b) False-color image of the inset in Fig.~\ref{fig:2}(c).
      (c) Schematic image of the actual electrode setup corresponding to (b).
        }
    \label{fig:G}
  \end{center}
\end{figure}

We initially tried to fabricate Corbino-geometry contacts, as schematically shown in Fig.~\ref{fig:G}(a). This configuration ensures the four-terminal measurement and the uniform current distribution. 
However, due to the small crystal size, the outer current electrode does not perfectly surround the central electrode (as schematically shown in Fig.~\ref{fig:G}(c)). 
This non-ideal geometry might induce the non-uniform current injection, although we perform standard four-terminal measurement. 
Nevertheless, we believe that our resistivity measurement is still valid based on the following two reasons.
First, the areas where the silver paste is attached can be regarded as equipotential surfaces. 
This is because the resistance of the silver paste is negligibly small compared to the in-plane resistance of top surface of the PtTe$_2$ crystal. 
Second, the measured out-of-plane resistivity is substantially larger than the in-plane resistivity. 
This high anisotropy assures that the measured resistance is dominated by the out-of-plane transport, even if the silver paste does not perfectly surround the contact.

\section{One-dimensional calculation\label{appendix:H}}

\begin{figure}
  \begin{center}
    \includegraphics[width = \linewidth]{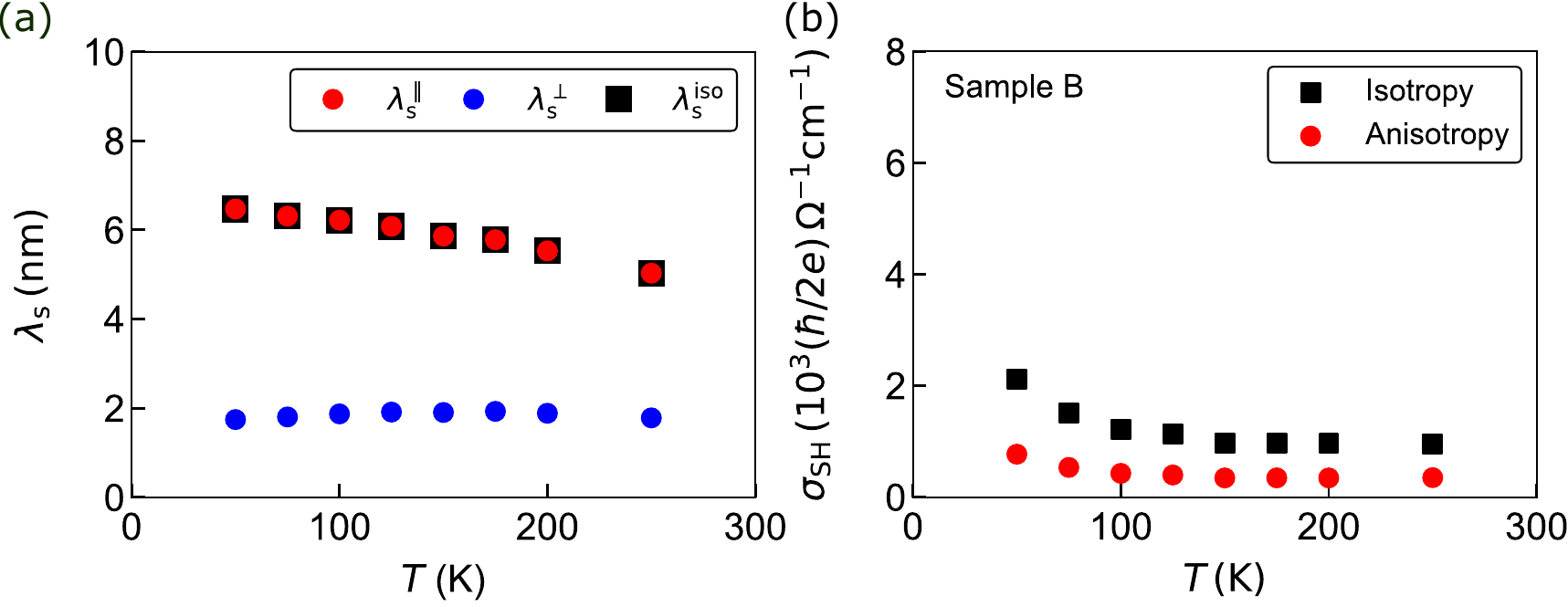}
    \caption{
      (a) Temperature dependence of spin diffusion lengths of PtTe$_2$ obtained from 1D calculations. 
      The black square shows the spin diffusion lengths $\lambda_{\mathrm{s}}^{\mathrm{iso}}$ obtained from the conventional isotropic analysis, assuming $\sigma = \sigma_{\parallel}$.
      The red and blue circles indicate the in-plane and out-of-plane spin diffusion lengths $\lambda_{\mathrm{s}}^{\parallel}$ and $\lambda_{\mathrm{s}}^{\perp}$, respectively, obtained from the anisotropic analysis. 
      (b) Temperature dependence of the spin Hall conductivity of PtTe$_2$ for Sample B. The red and black symbols represent calculated results from the anisotropic and isotropic analyses, respectively.
        }
    \label{fig:H}
  \end{center}
\end{figure}

We consider the 1D model of NLSV where the middle wire (M) is sandwiched by two ferromagnetic (F1 and F2) wires and bridged by a nonmagnetic (N) wire. 
As shown in Ref.~\cite{Niimi2012}, the 1D spin diffusion equation gives the following expressions for the ratio $\Delta R_{\mathrm{s}}^{\mathrm{with}}/\Delta R_{\mathrm{s}}^{\mathrm{without}}$ and the averaged injected spin current $\bar{I_{\mathrm{s}}}/I$ in middle wire,
\begin{widetext}
\begin{gather}
    \frac{\Delta R_{\mathrm{s}}^{\mathrm{with}}}{\Delta R_{\mathrm{s}}^{\mathrm{without}}}
    = \frac{2Q_{\mathrm{M}}\left\lbrace \sinh(L/\lambda_{\mathrm{N}}) + 2Q_{\mathrm{F}}(Q_{\mathrm{F}}+1)e^{L/\lambda_{\mathrm{N}}} \right\rbrace}
    {\cosh(L/\lambda_{\mathrm{N}})-(2Q_{\mathrm{F}}+1)\cosh((L-2d)/\lambda_{\mathrm{N}})+2Q_{\mathrm{M}}\sinh(L/\lambda_{\mathrm{N}}) + 2Q_{\mathrm{F}}(Q_{\mathrm{F}}+1) (2Q_{\mathrm{M}}+1)e^{L/\lambda_{\mathrm{N}}}} \label{eq:D1},
\end{gather}
\begin{align}
      \frac{\bar{I_{\mathrm{s}}}}{I} = & \frac{\lambda_{\mathrm{M}}}{t_{\mathrm{M}}}
    \frac{\cosh(t_{\mathrm{M}}/\lambda_{M})-1}{\sinh(t_{\mathrm{M}}/\lambda_{M})} \notag\\
    &\times\frac{p_{\mathrm{F}}Q_{\mathrm{F}}\left\lbrace(1+2Q_{\mathrm{F}})e^{(L-d)/\lambda_{\mathrm{N}}}-e^{-(L-d)/\lambda_{\mathrm{N}}}\right\rbrace}
    {\cosh{(L/\lambda_{\mathrm{N}})} -(2Q_{\mathrm{F}}+1)\cosh{((L-2d)/\lambda_{\mathrm{N}})}+ 2Q_{\mathrm{M}}\sinh(L/\lambda_{\mathrm{N}})+2Q_{\mathrm{F}}(Q_{\mathrm{F}}+1)(2Q_{\mathrm{M}}+1)e^{L/\lambda_{\mathrm{N}}}} \label{eq:D2},
\end{align}
\end{widetext}
where, for $ \mathrm{X} = \mathrm{N,\,F,\,M} $, we define the wire width as $ w_{\mathrm{X}} $, the thickness as $ t_{\mathrm{X}} $, the spin diffusion length as $ \lambda_{\mathrm{X}} $, 
$R_{\mathrm{X}}$ is the spin resistance, 
$Q_{\mathrm{M}}=R_{\mathrm{M}}/R_{\mathrm{N}}$, $Q_{\mathrm{F}}=R_{\mathrm{F}}/R_{\mathrm{N}}$, 
$p_{\mathrm{F}}$ is the spin polarization of F,
$d$ is the gap distance between the spin injector F1 and the middle wire M,
and $L$ is the gap distance between the spin injector F1 and the spin detector F2. 
The spin resistance $ R_{\mathrm{X}} $ is given by
$R_{\mathrm{F}} = \lambda_{\mathrm{F}}/\sigma_{\mathrm{F}} (1 - p_{\mathrm{F}}^2) A_{\mathrm{F}},\,
R_{\mathrm{N}} = \lambda_{\mathrm{N}}/\sigma_{\mathrm{N}} A_{\mathrm{N}},\,
R_{\mathrm{M}} = \lambda_{\mathrm{M}}/\sigma_{\mathrm{M}} A_{\mathrm{M}},$
where the cross-sectional areas are defined as $ A_{\mathrm{N}} = w_{\mathrm{N}} t_{\mathrm{N}} $, $ A_{\mathrm{F}} = w_{\mathrm{F}} w_{\mathrm{N}} $ and $A_{\mathrm{M}} = w_{\mathrm{N}} w_{\mathrm{M}} \tanh ( t_{\mathrm{M}}/\lambda_{\mathrm{M}})$.
From Eq.~\eqref{eq:D1}, the spin diffusion length can be determined.
This expression is applicable to conventional isotropic systems, but for anisotropic systems, the renormalization is required. To achieve this, the parameter $Q_{\mathrm{M}}$ needs to be rescaled as follows:
\begin{align}
    \quad&\sigma_{\mathrm{M}} = \sigma_{\parallel},  \quad t_{\mathrm{M}}'=\sqrt{\frac{\sigma_{\parallel}}{\sigma_{\perp}}}\,t_{\mathrm{M}}. \label{eq:D3}
\end{align}
The spin diffusion length obtained through this procedure is shown in Fig.~\ref{fig:H}(a).
Although the values from the 3D calculation tend to be slightly smaller than those from the 1D model, the order of magnitude and overall trends are consistent between the two approaches.
This indicates that the 1D model provides a sufficiently accurate approximation for the system.

The spin Hall conductivity can be calculated
from Eq.~\eqref{eq:D2} together with  
$\sigma_{\mathrm{SH}}=\sigma_{\mathrm{M}}^2w_{\mathrm{M}}\Delta R_{\mathrm{ISHE}}/x(\bar{I_{\mathrm{s}}}/I)$, where $x$ is the shunting coefficient of middle wire by nonmagnetic wire~\cite{Sagasta2016,Isasa2015}.
Although the shunting effect is inherently included in 3D calculations, it cannot be captured by 1D calculations.
The shunting coefficient is typically obtained experimentally or via 3D simulations. Since the experimental determination was not feasible in our case, we employed 3D simulations.
By applying the coordinate transformation given by Eq.~\eqref{eq:D3}, this expression can be reformulated in the same framework as developed in Sec.~\ref{sec:2}. The spin Hall conductivity then becomes
\begin{align}
    \sigma_{\mathrm{SH}}
    &=\sqrt{\frac{\sigma_{\perp}}{\sigma_{\parallel}}}\sigma_{\mathrm{SH}}'
    =\sqrt{\frac{\sigma_{\perp}}{\sigma_{\parallel}}}\sigma_{\parallel}^2\frac{w_{\mathrm{M}}}{x'}\left(\frac{I}{\bar{I_{\mathrm{s}}}'}\right)\Delta R_{\mathrm{ISHE}},
\end{align}
where $\bar{I_{\mathrm{s}}}'/I$ and $x'$ are calculated using Eq.~\eqref{eq:D3} and $\lambda_{\mathrm{M}} = \lambda_{\mathrm{s}}^{\parallel}$.
Figure~\ref{fig:H}(b) shows the temperature dependence of the spin Hall conductivity obtained from anisotropic ($\sigma_{\mathrm{SH}}$) and isotropic ($\sigma_{\mathrm{SH}}^{\mathrm{iso}}$) analyses. 
Just like in 3D calculations, the results satisfy the relation $\sigma_{\mathrm{SH}} < \sigma_{\mathrm{SH}}^{\mathrm{iso}}$. 
However, the obtained values are smaller than those from 3D calculations.

\section{Computational conditions for first-principles calculations\label{appendix:I}}

First-principles calculations based on density functional theory were performed using the Perdew-Burke-Ernzerhof parameterization of the generalized gradient approximation~\cite{Perdew1996} and the projector-augmented wave (PAW)~\cite{Blochl1994} method as implemented in Quantum ESPRESSO~\cite{Giannozzi2009, Giannozzi2017, Giannozzi2020}.
We used PAW pseudopotentials provided in pslibrary~\cite{Corso2014}.
SOI was included in our calculations.
The plane-wave cutoff energies for Kohn--Sham orbitals and charge density are 60 and 800 Ry, respectively.
Spin Hall conductivity was calculated using Wannier orbitals~\cite{Qiao2018} as implemented in Wannier90~\cite{Pizzi2020}.

First, we calculated the spin Hall conductivity for bulk Pt. The lattice constant of face-centered-cubic Pt ($a= 3.9231$ \AA ) was taken from Ref.~\onlinecite{Wyckoff1963}.
After self-consistent-field (SCF) calculation using a $16\times 16\times 16$ ${\bm k}$-mesh, we constructed the Wannier orbitals of Pt-$s/p/d$ orbitals using the same ${\bm k}$-mesh.
For Wannierization, outer and inner energy windows were set as [$-18$:42] and [$-18$:12] eV, respectively, where the Fermi energy was set to zero.
Spin Hall conductivity was evaluated using a $100\times 100\times 100$ ${\bm k}$-mesh with $3\times 3\times 3$ adaptive refinement with the refinement threshold of 100 \AA$^2$.
We got $\sigma^{y}_{zx} = 4.4 \times 10^3\,(\hbar/2e)\mathrm{\Omega^{-1}cm^{-1}}$, as consistent with previous theoretical studies~\cite{Guo2008,Qiao2018}.

For PtTe$_2$, we used the experimental crystal structure ($a=4.0259$ \AA, $c=5.2209$ \AA, $z=0.254$ with the space group $P\bar{3}m1$) taken from Ref.~\onlinecite{Furuseth1965}.
SCF calculation and Wannierization of Pt-$s/p/d$ and Te-$s/p/d$ orbitals were performed using a $16\times 16\times 12$ ${\bm k}$-mesh.
For Wannierization, outer and inner energy windows were set as [$-20$:40] and [$-20$:7.3] eV, respectively, where the Fermi energy was set to zero.
Spin Hall conductivity was evaluated using a $72\times 72\times 60$ ${\bm k}$-mesh with $3\times 3\times 3$ adaptive refinement with the refinement threshold of 100 \AA$^2$.

\end{document}